\documentclass[twocolumn]{aastex631}
\usepackage[utf8]{inputenc}
\usepackage{amsmath}
\usepackage{natbib}
\usepackage{bm}
\usepackage{multirow}

\newcommand{\p}{\partial}

\newcommand{\ii}{\mathrm{i}}
\newcommand{\C}{\mathcal{C}^\lambda}

\newcommand{\avg}[1]{\langle{#1}\rangle}

\newcommand{\dd}{\delta}

\newcommand{\Hgas}{H_\mathrm{g}}

\newcommand{\hgas}{h_\mathrm{g}}

\newcommand{\rhod}{\rho_\mathrm{d}}
\newcommand{\rhog}{\rho_\mathrm{g}}

\newcommand{\sigg}{\Sigma_\mathrm{g}}

\newcommand{\taus}{\tau_\mathrm{s}}
\newcommand{\st}{\mathrm{St}}
\newcommand{\w}{ \widetilde{W}}

\newcommand{\Hg}{H_\mathrm{g}}
\newcommand{\Hd}{H_\mathrm{d}}

\newcommand{\OmK}{\Omega_{\rm K}}
\newcommand{\OmKref}{\Omega_0}
\newcommand{\etatot}{\eta_\mathrm{tot}}

\newcommand{\etatilde}{\widetilde{\eta}}

\newcommand{\alphass}{\alpha_\mathrm{SS}}
\newcommand{\rhodmax}{\rho_\mathrm{d,max}}
\newcommand{\rhoroche}{\rho_\mathrm{R}}
\newcommand{\Wpert}{\frac{\dd\rhog}{\rhog}}
\newcommand{\Qpert}{\frac{\dd\rhod}{\rhod}}

\defcitealias{MK22}{LH22}

\shorttitle{Streaming instabilities in accreting disks}
\shortauthors{C.-Y. Hsu and M.-K. Lin}

\begin{document}


\title{Nonlinear evolution of streaming instabilities in accreting protoplanetary disks} 

\correspondingauthor{Min-Kai Lin}
\email{mklin@asiaa.sinica.edu.tw}

\author{Chun-Yen Hsu}
\altaffiliation{Current address: Department of Astronomy, University of Virginia, Charlottesville, VA 22904, USA}
\affiliation{Institute of Astronomy and Astrophysics, Academia Sinica, Taipei 10617, Taiwan}

\author[0000-0002-8597-4386]{Min-Kai Lin}
\affiliation{Institute of Astronomy and Astrophysics, Academia Sinica, Taipei 10617, Taiwan}
\affiliation{Physics Division, National Center for Theoretical Sciences, Taipei 10617, Taiwan}

\begin{abstract}
The streaming instability (SI) is one of the most promising candidates for triggering planetesimal formation by producing dense dust clumps that undergo gravitational collapse. Understanding how the SI operates in realistic protoplanetary disks (PPDs) is therefore crucial to assess the efficiency of planetesimal formation. Modern models of PPDs show that large-scale magnetic torques or winds can drive laminar gas accretion near the disk midplane. In a previous study, we identified a new linear dust-gas instability, the azimuthal drift SI (AdSI), applicable to such accreting disks and is powered by the relative azimuthal motion between dust and gas that results from the gas being torqued. 
In this work, we present the first nonlinear simulations of the AdSI. 
We show that it can destabilize an accreting, dusty disk even in the absence of a global radial pressure gradient, which is unlike the classic SI. We find the AdSI drives turbulence and the formation of vertically-extended dust filaments that undergo merging. In dust-rich disks, merged AdSI filaments reach maximum dust-to-gas ratios exceeding 100. Moreover, we find that even in dust-poor disks the AdSI can increase local dust densities by two orders of magnitude. We discuss the possible role of the AdSI in planetesimal formation, especially in regions of an accreting PPD with vanishing radial pressure gradients. 
\end{abstract}

\section{Introduction}\label{intro}




A key step in the core accretion scenario of planet formation is the formation of 1---100 km or larger-sized planetesimals \citep{chiang10, johansen17, raymond22, drazkowska22}. Planetesimal formation is often attributed to the self-gravitational collapse of dust grains or pebbles. A necessary condition for this to occur is for solids to be concentrated to a sufficiently high volume density relative to the ambient gas in protoplanetary disks (PPDs) \citep{goldreich73, youdin02, shi13, gerbig20}. To this end, several dust concentration mechanisms can operate in PPDs, e.g. vertical settling, zonal flows or pressure bumps, vortices, etc. \citep{johansen14,pinilla17}, and the \emph{streaming instability} \citep[SI,][]{youdin05,youdin07,johansen07}.

Among these, the SI has perhaps garnered the most attention \citep[e.g.][see also the recent review by \citealt{lesur22}]{bai10, yang14, carrera15, yang17, flock21, li21}. When the dust surface density divided by the gas surface density exceeds a critical value, the SI can produce dust clumps that subsequently undergo gravitational collapse into planetesimals \citep{johansen09,johansen11,simon17,schafer17,li19}. However, the SI itself does not require self-gravity.

The SI is powered by the relative motion between dust and gas in PPDs. Usually, this arises from the fact that the gas rotation is slightly sub-Keplerian due to a (negative) radial pressure gradient, but solids tend to rotate at the full Keplerian speed. The resulting headwind on the solids causes it to lose angular momentum to the gas and drift inwards \citep{whipple72,weiden77}, while the gas drifts outward. This relative dust-gas drift provides the free energy for instability. However, the precise mechanism for the SI is rather subtle and a variety of interpretations have been developed. These include dust trapping by pressure maxima, pressure-density phase lags, or resonances between waves in the gas and dust-gas drift \citep{jacquet11, lin17, squire18b, squire20, pan20b}. 

Recent extensions of the SI have begun to incorporate additional effects to better understand how it operates under more general disk conditions, e.g. turbulence \citep{gole20,schafer20}, vertical disk stratification \citep{lin21}, or radial disk structures \citep{carrera21a,carrera22}. It is worth noting that observations of PPDs indeed show that dust rings appear commonplace, which likely reflect a non-monotonic radial gas distribution \citep{dullemond18, andrews20}. 

At the same time, the latest theoretical models of PPDs show that their overall gas dynamics are controlled by large-scale magnetic winds and torques \citep[see reviews by][and references therein]{lesur20,pascucci22}. 
Of particular relevance here are models that exhibit non-turbulent gas accretion around the disk midplane   \citep[e.g.][]{bai17,bethune17,wang19,gressel20,cui21b}, where pebbles are expected to settle and form planetesimals. A natural question is then how does the SI operate in such accreting disks? 

Furthermore, modern simulations often show that PPDs can spontaneously develop axisymmetric pressure bumps \citep{bethune16,suriano18, hu19}, which efficiently trap dust \citep{krapp18,riols20}. In addition to possibly explaining sub-structures in observed disks, the resulting dust rings may be also be preferential sites for the SI, since it grows on dynamical timescales when the local dust-to-gas ratio is greater than of order unity \citep{chen20}, which can be expected in a dust-trapping pressure bump. However, a complication is that as one approaches a pressure maximum, the SI's characteristic lengthscale becomes arbitrarily small. At the exact bump center, the instability ceases altogether because there is no radial pressure gradient and hence no radial drift between dust and gas. 

Motivated by the above considerations, in a previous study we generalized the linear theory of the SI to account for a background gas accretion flow and considered a range of radial pressure gradients, including zero \citep[][hereafter \citetalias{MK22}]{MK22}. We discovered a new form of the SI powered by the \emph{azimuthal} velocity difference between dust and gas, which is ultimately driven by the (magnetic) torque that mediates gas accretion. This `azimuthal drift' SI (AdSI) operates even in the absence of a radial pressure gradient, suggesting it could be relevant in regions near a pressure bump. 

In this work, we present numerical simulations of the AdSI. Our main goal is to confirm its existence and to compare its nonlinear evolution with the \emph{classical} SI of \citeauthor{youdin05}. We find the AdSI can indeed develop and drive turbulence without a background radial pressure gradient. In strongly accreting disks, it can produce dust-to-gas ratios for which gravitational collapse is expected. Moreover, the AdSI can lead to appreciable dust concentrations, even when the initial dust-to-gas ratio is below unity, which is unlike the classic SI. 


This paper is organized as follows. In \S\ref{Disk_model} we describe our local disk model and basic equations. Our numerical methodology, including simulation diagnostics, are detailed in \S\ref{Numerical_method}. We present simulation results in \S\ref{Non_linear}, including analyses of turbulence, dust drift, and dust concentrations. We discuss our findings in the context of planetesimal formation in \S\ref{discussion} and summarize in \S\ref{summary}.

\section{Disk Model} \label{Disk_model}

We consider a PPD of gas and dust orbiting a star of mass $M_*$. Cylindrical co-ordinates $(R,\phi, z)$ are centered on the star. We assume an isothermal gas with a constant sound-speed $C_s = \Hgas\OmK$, where $\Hgas$ is the pressure scale height, $\OmK(R)=\sqrt{GM_*/R^3}$ is the Keplerian frequency, and $G$ is the gravitational constant. 

The disk is threaded by a magnetic field that is assumed to remain passive, i.e. it does not respond to the gas dynamics, which might be expected for weakly ionized gas in PPDs \citep{lesur20}. The magnetic field, however, drives gas accretion onto the star through horizontal Maxwell stresses or by extracting angular momentum vertically \citep{bai16,lesur21,tabone22}. We realize this accretion flow in a hydrodynamic model by applying an external torque onto the gas. See \cite{mcnally17} for a similar approach for simulating planets interacting with accreting disks.

We include a single species of uncharged dust grains with a stopping time $\taus$ that characterizes the frictional drag with the gas. We consider small grains with Stokes numbers $\st \equiv \taus \OmK \ll 1$, which are tightly -- though not necessarily perfectly -- coupled to the gas. In this limit, one can treat the dust population as a pressureless fluid \citep{jacquet11}.

\subsection{Governing equations} \label{Governing_Eq}

We focus on a small patch of the disk around a fiducial point $(R_0,\phi_0,0)$ with $\phi_0(t)=\OmKref t $, where $\OmKref\equiv\OmK(R_0)$, $t$ is the time, and adopt the shearing box framework \citep{goldreich65} with Cartesian coordinates $(x,y,z)$ corresponding to the radial, azimuthal, and vertical directions in the global disk. For a small box and lengthscales of interest $\ll R_0$, we can ignore curvature effects and approximate Keplerian rotation as the linear shear flow $\bm{U}_\mathrm{K}=-\frac{3}{2} x\OmKref \hat{\bm{y}}$. We consider dynamics close to disk midplane and neglect the vertical component of stellar gravity. We assume axisymmetry throughout so that  $\p_y \equiv 0$. The total gravitational and centrifugal force in the box is $3 x \OmKref^2 \hat{\bm{x}}$. For clarity, we henceforth drop the subscript zero that denotes the evaluation of global quantities at the reference radius, which includes $\Hg$.  

The axisymmetric, unstratified shearing box equations for our dusty-gas disk are 
\begin{align}
&\frac{\partial \rhog }{\partial t} + \nabla \cdot (\rhog \bm{v}) = 0 \label{gas_mass_local},\\
&\frac{\partial \bm{v} }{\partial t} + \bm{v}\cdot \nabla \bm{v} = 2 v_y \Omega \hat{\bm{x}} - v_x \frac{\Omega}{2} \hat{\bm{y}} - \frac{1}{\rhog} \nabla P \notag \\ 
&\phantom{\frac{\partial \bm{v} }{\partial t} + \bm{v}\cdot \nabla \bm{v} =}
 + 2\etatot R\Omega^2\hat{\bm{x}} + F_{\phi}\hat{\bm{y}} \notag\\
 & \phantom{\frac{\partial \bm{v} }{\partial t} + \bm{v}\cdot \nabla \bm{v} =}
  + \frac{\epsilon}{\taus}(\bm{ w}-\bm{v}) + \frac{1}{\rhog}\nabla\cdot\bm{T},\label{gas_mom_local}\\
&\frac{\partial \rho_d }{\partial t} + \nabla \cdot (\rho_d \bm{w}) = \nabla\cdot\left(D\rhog\nabla\epsilon\right), \label{dust_mass_local}\\
&\frac{\partial \bm{w} }{\partial t} + \bm{w} \cdot \nabla \bm{w} = 2 w_y \Omega \hat{\bm{x}} - w_x \frac{\Omega}{2} \hat{\bm{y}} - \frac{1}{\taus}(\bm{w}-\bm{v})   \label{dust_mom_local}
\end{align}
\citepalias{MK22}, where $\rhog, \rhod$ are the (midplane) gas and dust densities, and $\bm{v},\,\bm{w}$ are their velocities relative to the Keplerian shear flow, respectively. We also define $\epsilon \equiv \rhod/\rhog$ as the dust-to-gas ratio. We assume $\taus$, or equivalently $\st$, is constant in the box. Note that here $P=C_s^2\rhog$ is the { local} pressure fluctuation and is zero in equilibrium.

In the gas momentum equation (\ref{gas_mom_local}), we model two effects from the global disk as body forces in the local box. The term $\propto \etatot$ represents the combined global gas and magnetic pressure radial gradients. In the usual case of weak fields and a negative gas pressure gradient, this leads to a sub-Keplerian gas flow. Such a constant radial forcing is commonly used to model the SI in the local approximation \citep[e.g.][]{johansen07}. 

The forcing $F_{\phi}<0$ represents { the azimuthal Lorentz force from the global magnetic field, which exerts a torque on the gas and drives accretion.} Note that because we assume the magnetic field is passive, no induction equation is needed. 

{ In the local approach, $\etatot$ and $F_\phi$ are taken to be constant and independent input parameters. However, in a resistive disk threaded by a spiral magnetic field, these are related to the global disk profiles as  
\begin{align}
    &\eta_\mathrm{tot} = \eta + \frac{F_R}{2R\OmK^2}\label{etatot},\\
    &F_\phi = \frac{B_R B_\phi}{2\mu_0R\rhog}\label{lorentz_force},
\end{align}
where 
\begin{align}
 \eta = -\frac{1}{2R\OmK^2\rhog}\frac{\p P_\mathrm{glb}}{\p R}
\end{align}
is the dimensionless radial gas pressure gradient and $F_{R}$ is the radial component of the Lorentz force, 
\begin{align}
     &F_R = -\frac{B_\phi^2}{2\mu_0R\rhog}.
\end{align}
In the above expressions, $P_\mathrm{glb}$ is the global pressure distribution, $B_{R,\phi}$ are the radial and azimuthal components of the magnetic field (see \citetalias{MK22} for explicit expressions), respectively, and $\mu_0$ is the magnetic permeability. Note that $B_RB_\phi<0$. 
}

For completeness, we also include a viscous stress tensor $\bm{T}$ in the gas momentum equation (\ref{gas_mom_local}), which is given by
\begin{align}
    \bm{T} = \rhog\nu \left(\nabla\bm{v} + \nabla\bm{v}^\dagger - \frac{2}{3}\bm{I}\nabla\cdot\bm{v}\right),\label{viscous_stress}
\end{align}
with a constant kinematic viscosity $\nu$. We use gas viscosity as a proxy for any underlying turbulence. Particle-stirring by said turbulence is then modeled as the diffusion term $\propto D$ in the dust mass equation { (\ref{dust_mass_local})}. However, for the most part, we neglect viscosity and diffusion except in code tests and \S\ref{addsims}. When  considered, we set $D = \nu = \alpha_\mathrm{visc} C_s\Hgas$, where $\alpha_\mathrm{visc}$ is a constant parameter.  

\subsection{Physical parameters} \label{Physical_para}

This subsection describes all of the physical parameters that characterize our models. The disk aspect-ratio is 
\begin{align}
\hgas \equiv \frac{\Hgas}{R}, 
\end{align}
and we take $\hgas=0.05$ in all computations. We also define a reduced  pressure gradient parameter 
\begin{align}
    \etatilde \equiv \frac{\eta_\mathrm{tot}}{\hgas}. 
\end{align}
Typically $\etatilde$ is of $O(\hgas)$ in PPDs, but we will vary $\etatilde$ to explore how the SI behaves with vanishing pressure gradients,  $\etatilde\to0$.

We also define the dimensionless azimuthal forcing 
\begin{align}
    \alpha_M = -\frac{2R F_\phi}{C_s^2}, \label{alphaMdef}
\end{align}
which can be related to horizontal Maxwell stresses if the torque results from a spiral magnetic field in a resistive disk \citepalias[e.g.][]{MK22}. While we are motivated by accretion mediated by large-scale magnetic fields, our results are also applicable to accretion driven by other means, as long as it can be represented by an $F_\phi$ in the gas' azimuthal equation of motion. Nevertheless, we will refer to $\alpha_M$ as the Maxwell stress for convenience. 

Each of our disk models is characterized by $\st$, $\etatilde$, $\alpha_M$, and the initial value of $\epsilon$. However, to limit the volume of parameter space for computational feasibility, we fix $\st = 0.1$ throughout this paper. This corresponds to cm-sized grains with internal density $1$ g cm$^{-3}$ at $20$ au in a Minimum Mass Solar Nebula-like disk \citep{chiang10}. 

\subsection{Equilibrium state}\label{eqm_state} \label{Initial_cond}
We consider steady-state solutions of Eqs. \ref{gas_mass_local}--\ref{dust_mom_local} with constant $\rhod,\,\rhog$ and velocity deviations given by: 
 \begin{align}
& \frac{v_{x}}{C_s} = \frac{2 \epsilon \st}{\Delta^2} \etatilde - \frac{\alpha_M\hgas\left(\st^2 + \epsilon +1\right)}{\Delta^2}, \label{eqm_vr}\\
& \frac{v_{y}}{C_s} = - \frac{\left(\st^2 + \epsilon +1\right)}{\Delta^2} \etatilde - \frac{\alpha_M\hgas\epsilon \st}{2\Delta^2}, \label{eqm_vphi}\\
& \frac{w_{x}}{C_s} = - \frac{2 \st}{\Delta^2}\etatilde - \frac{\alpha_M\hgas (\epsilon +1)}{\Delta^2},  \label{eqm_wr}  \\
& \frac{w_{y}}{C_s} = - \frac{\left(\epsilon +1\right)}{\Delta^2}\etatilde + \frac{\alpha_M\hgas\st}{2\Delta^2}\label{eqm_wphi},
\end{align}
where $\Delta^2 \equiv \st^2 + (1 + \epsilon)^2$. The first and second terms on the RHS correspond to drift induced by the large-scale pressure gradient and magnetic torque, respectively. Note that pressure gradients dominate the radial drift between dust and gas; while the magnetic torque dominates their azimuthal drift \citepalias{MK22}. Note that viscosity and diffusion do not affect these equilibrium solutions. See \cite{carrera21b} for a generalization that includes a full pressure bump in the box. 

\section{Numerical Method} \label{Numerical_method}


We adapt \textsc{fargo3d} \citep{llambay16} with its multi-fluid extension \citep{llambay19} to evolve the dusty shearing box equations \ref{gas_mass_local}--\ref{dust_mom_local}. \textsc{fargo3d} is a versatile finite-difference code but is particularly suited for disk problems. 

We made two augmentations to the shearing box module in the public version of \textsc{fargo3d}. First, as the original code solves the full velocity field (e.g., $\bm{v}+\bm{U}_\mathrm{K}$), we subtract the background shear flow from the outset by removing the centrifugal source term ($3\Omega^2x$) that would appear in the equations for the full $x$-velocities. We also treat the second term on the right-hand side of the $y$-momentum equations (\ref{gas_mom_local} and \ref{dust_mom_local}) explicitly in the source step, as opposed to absorbing it in the transport step as done for the $y$-Coriolis force in the original code. Our approach is similar to that done in the \textsc{athena} code \citep{stone10}. Second, we added a constant azimuthal forcing $F_\phi$ in the gas $v_y$ equation.

{ For consistency with the dust diffusion term adopted in Eq. \ref{dust_mass_local}, which the linear theory developed in \citetalias{MK22} is based upon, we}
also modified \textsc{fargo3d}'s dust diffusion module such that the diffusive mass flux is proportional to $\nabla \left(\rhod/\rhog\right)$, rather than $\nabla\left[\rhod/\left(\rhod+\rhog\right)\right]$ as in the standard release. 

{ We enable the `FARGO' algorithm \citep{masset00a, masset00b}, originally designed to speed up the simulations by decomposing the total flow velocity into the average orbital motion and residuals during the advection step.  However, since we remove the background Keplerian flow from the outset, this choice makes little difference.}

{ In Appendix \ref{code_test}, we test the revised code against the linear theory of the AdSI as described in \citetalias{MK22}.}

\subsection{Simulation setup} 

Our simulations are three-dimensional but axisymmetric, or `2.5D'. { In practice, this is realized by setting the azimuthal ($y$) grid to one cell wide.} The { meridional} domain is $(x,z)\in[-L_{x,z}/2, L_{x,z}/2]$ with $L_x=0.2\Hgas$ and $L_z=0.05\Hgas$. The small vertical domain is chosen for consistency with our unstratified approximation that focuses on the disk midplane. We use $N_x=2048$ and $N_z=512$ cells in the radial and vertical directions, respectively, which gives a resolution of about $10^{-4}\Hgas$. This resolution was chosen as a compromise between capturing as wide of a range of AdSI modes as possible, since it can develop on arbitrarily small scales in inviscid disks \citepalias{MK22}, and the computational cost. The same applies to the classic SI: as $\etatilde\to0$ its  characteristic lengthscale (in units of $\Hg$) { vanishes}. 


We apply strictly periodic boundary conditions to both radial and vertical directions and run most simulations to $t = 50P$, where $P = 2 \pi/ \Omega $ is orbit period. We use a Courant–Friedrichs–Lewy (CFL) number of $0.15$. Finally, we adopt units such that $C_s = H_g = \Omega = 1$. For non-self-gravitating disks, the density scale is arbitrary, we thus define the equilibrium gas density $\rho_0 = 1$ for convenience. 

\subsection{Main runs}

Table \ref{table1} lists our main simulations. We investigate two classes of disks: dust-rich ($\epsilon = 3$) and dust-poor ($\epsilon = 0.2$). We consider  $\etatilde\in[0,0.005,0.05]$ to mimic regions at a pressure bump, weak pressure gradients, and typical pressure gradients, respectively; and $\alpha_M\in[0,0.01, 0.1]$ to represent varying degrees of underlying gas accretion. Runs are labeled by the above parameters, e.g. E3eta005am0 corresponds to $(\epsilon, \etatilde,\alpha_M) = (3, 0.05, 0)$. 

In Table \ref{table1}, we also describe the end state of each run as: `stable' if no instability develops; `unsaturated' if the instability grows but does not saturate within the simulation timescale; `turbulent' if the system saturates but does not produce strong clumping; and `clumping' if the system is turbulent and produces strong clumping. The clumping condition is defined in \S\ref{clumping_cond}.

\begin{table}
\caption {Parameters for our main simulations. \label{table1} 
} 
\begin{ruledtabular}
\begin{tabular}{llll}
 Name  & $\etatilde$ &  $\alpha_M$ & End state \\
        \hline
        \multicolumn{4}{c}{$\epsilon = 3$, $N_p = 2^{26}$} \\
        \hline
 E3eta005am0      & $0.05$  & $0$   & Clumping  \\
 E3eta005am001    & $0.05$  & $0.01$& Clumping  \\
 E3eta005am01     & $0.05$  & $0.1$ & Clumping  \\
 
 E3eta0005am0     & $0.005$ & $0$   & Turbulent \\
 E3eta0005am001   & $0.005$ & $0.01$& Clumping  \\
 E3eta0005am01    & $0.005$ & $0.1$ & Clumping  \\
 
 E3eta0am0        & $0$     & $0$   & Stable    \\
 E3eta0am001      & $0$     & $0.01$& Turbulent \\
 E3eta0am01       & $0$     & $0.1$ & Clumping  \\
       \hline
        \multicolumn{4}{c}{$\epsilon = 0.2$, $N_p = 2^{22}$} \\
        \hline
 E02eta005am0     & $0.05$  & $0$   & Unsaturated \\
 E02eta005am001   & $0.05$  & $0.01$& Unsaturated \\
 E02eta005am01    & $0.05$  & $0.1$ & Unsaturated \\
 
 E02eta0005am0    & $0.005$ & $0$   & Unsaturated \\
 E02eta0005am001  & $0.005$ & $0.01$& Unsaturated \\
 E02eta0005am01   & $0.005$ & $0.1$ & Turbulent \\
 
 E02eta0am0       & $0$     & $0$   & Stable    \\
 E02eta0am001     & $0$     & $0.01$& Turbulent \\
 E02eta0am01.     & $0$     & $0.1$ & Turbulent \\
\end{tabular}
\end{ruledtabular}
\end{table}

\subsection{Initial conditions and perturbations}

The disk is initialized with the equilibrium solutions described in \S\ref{Initial_cond}. To trigger instability, we perturb the initial dust density field by adding `particles' to the disk as follows. We randomly select $N_p$ points in the domain and assign $\pm10^{-4}\rho_0$ to each point, again at random. For each grid cell that contains $n_p$ points or particles, the cell's dust density perturbation is then $10^{-4}N_p\rho_0$. The total dust density perturbation over the domain is close to zero. 

\subsection{Diagnostics} \label{Diagnostics}

In this subsection, we describe the methods adopted for analyzing the turbulence properties and assessing dust clumping in our simulations.

\subsubsection{Transport and turbulence}

We follow \cite{johansen07} and define 
\begin{align}
    & \alpha_{\rm SS} \equiv \frac{ \rho_g v_x v_y }{ \rho_0 C_s^2} , \label{gas_stress}
\end{align}
as a dimensionless radial angular momentum flux carried by the gas. 
The numerator and denominator of $\alpha_{\rm SS}$ are the Reynolds stress and (equilibrium) thermal pressure, respectively. This quantity is similar to the Shakura–Sunyaev stress parameter \citep{shakura73} used by \cite{yang18} and \cite{xu21} in their particle-gas simulations. Note that $\alpha_\mathrm{SS}$ includes a laminar contribution from the equilibrium gas velocity field (Eqs. \ref{eqm_vr}--\ref{eqm_vphi}).

Using $\avg{\cdot}$ to denote averaging over the $x-z$ plane, we first calculate $\langle \alpha_{\rm SS} \rangle$, then further conduct a time average as 
\begin{align}
& \overline{\langle \alpha_{\rm SS} \rangle} \equiv  \frac{1}{t_2-t_1} \int_{t_1}^{t_2}  \langle \alpha_{\rm SS} \rangle dt . \label{gas_stress_time_aver}
\end{align}
Based on the time at which our simulations reach saturation, we use $t_1 = 40P$ and $t_2 = 50P$. For this interval, we output the simulation data every $0.01P$ and perform the time integration explicitly. 


The next quantity of interest is the bulk gas diffusion coefficient $D_{g,i}$ in the $i^\mathrm{th}$ direction \citep{yang18}. We define its dimensionless equivalent as 
\begin{align}
& \alpha_{g,i} \equiv \frac{D_{g,i}}{C_s H_g} \simeq  \left( \frac{\overline{\delta v_{i}}}{C_s} \right)^2 \tau_{c,i} , \label{gas_diffusion_coef}
\end{align}
where
\begin{align}
& \delta v_{i} \equiv \sqrt{\avg{v_{i}^2} - \avg{v_{i}}^2}, \label{fluc_vel} 
\end{align}
 is the dispersion in the $i^\mathrm{th}$ velocity component with its time average $\overline{\delta v_i}$ defined in a similar manner to Eq. \ref{gas_stress_time_aver}; 
and 
\begin{align}
& \tau_{c,i} \equiv \Omega t_{c,i}, \label{dimensionless_cor_time}
\end{align}
is a dimensionless measure of the correlation time $t_{c,i}$. 

We measure $t_{c,i}$ by plotting the auto-correlation function of the $i^\mathrm{th}$ velocity component,
\begin{align}
& R_i(t) \equiv \int_{t_1}^{t_1+5P} [v_i(\tau) - \overline{v_i}][v_i(\tau+t) - \overline{v_i}] d\tau , \label{gas_autorelation}
\end{align}
where
\begin{align}
& \overline{v_i} \equiv \frac{1}{t_2-t_1} \int_{t_1}^{t_2} v_{i}(\tau^\prime) d\tau^\prime
\label{mean_vel}
\end{align}
is the mean gas velocity, and 
define $t_{c,i}$ as half-life of $R_i$. We calculate $R_i(t)$ for each cell  and use $\langle R_i(t) \rangle$ to obtain an averaged auto-correlation function. An example of this procedure is given in \S\ref{turbulence}. 

We also examine the gas' turbulent spectra by first computing the vertically-averaged kinetic energy density $\avg{\rhog v^2}_z$, which is a function of $x$ and time. We then take its Fourier transform in $x$, which gives the amplitude of modes with radial wavenumber $k_x$. We scale the wavenumber by $\Hg$ and thus plot the Fourier modes $\widehat{\avg{\rhog v^2}_z}$ as a function of $K_x\equiv k_x\Hg$. Note that here and below $\avg{\cdot}_z$ denotes a vertical average.

\subsubsection{Clumping condition}\label{clumping_cond}
One of the main goals of this paper is to assess whether or not a given system will lead to planetesimal formation. Since we do not include self-gravity, we follow other authors  \citep[e.g.][]{li21,xu21} and measure the maximum dust density $\rho_\mathrm{d,max}$ and compare it to the Roche density, 
\begin{align}
& \rho_R \equiv \frac{9\Omega^2}{4 \pi G}. \label{Roche density}
\end{align}
A dust clump with $\rhodmax>\rho_\mathrm{R}$ can be expected to undergo gravitational collapse \citep[but see][ for a more stringent criterion in the case of perfectly-coupled dust]{shi13}, provided it can overcome internal dust diffusion \citep{klahr20}. In this work, for simplicity, we only consider the Roche density criterion, which should be taken as a necessary but not sufficient condition. Note that simulating gravitational collapse requires modeling the dust as Lagrangian particles, rather than the fluid approach taken here. 

One can calculate $\rhoroche/\rhog$ for a given disk model with Toomre parameter $Q\equiv C_s\Omega/\pi G \sigg$, where the gas surface density $\sigg = \sqrt{2\pi}\rhog$, which gives $\rhoroche \simeq 5.6Q\rhog$. For example, \cite{li21} consider a low-mass disk with $Q = 32 $ and define $\rhoroche \simeq 180\rhog$; while \cite{xu21} consider a minimum-mass solar nebula disk with $\rhoroche \sim 130 \rhog $ at 30 AU and $\rhoroche\sim 300\rhog$ at 1AU. For convenience,  we define strong clumping as 
\begin{align}
    \frac{\rhodmax}{\rho_0} > 100\quad \text{(strong clumping)},\label{strong_clump_cond}
\end{align}
which would lead to gravitational collapse for $Q\lesssim 18$.

\section{Results}\label{Non_linear}

\begin{figure}
    \centering
    \includegraphics[width=\linewidth]{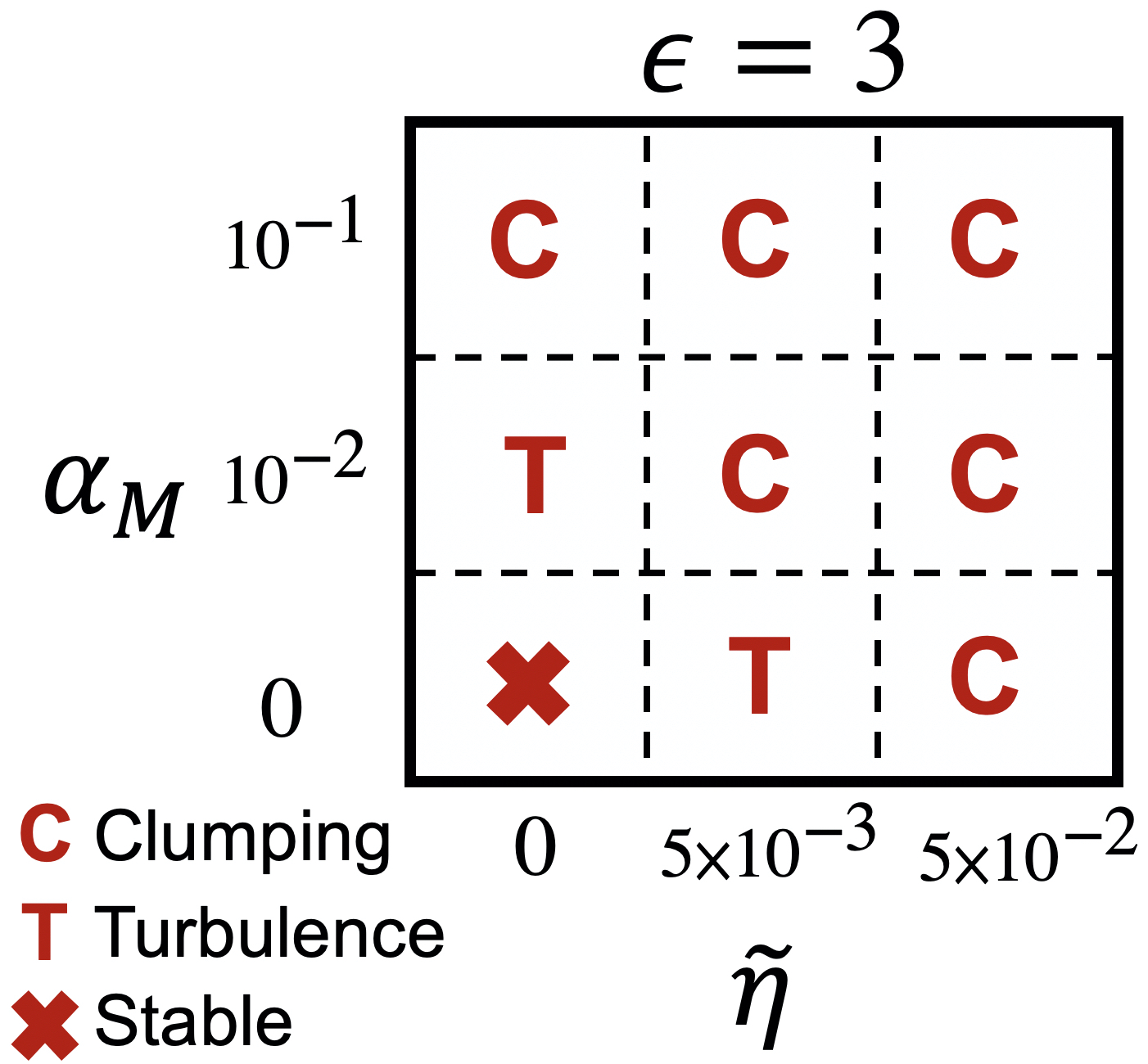}
    \caption{
   The outcome of our main simulation with $\epsilon=3$ for different Maxwell stresses, $\alpha_M$, and global radial pressure gradients, $\etatilde$. Turbulent cases that meet the clumping criterion (Eq. \ref{strong_clump_cond}) are symbolized by `C'. Cases that are turbulent but do not meet the clumping condition are marked by `T'. The case with $\etatilde=\alpha_M=0$ remains stable. 
    }
    \label{Paper_overview}
\end{figure}
\begin{figure}
    \centering
    \includegraphics[width=\linewidth]{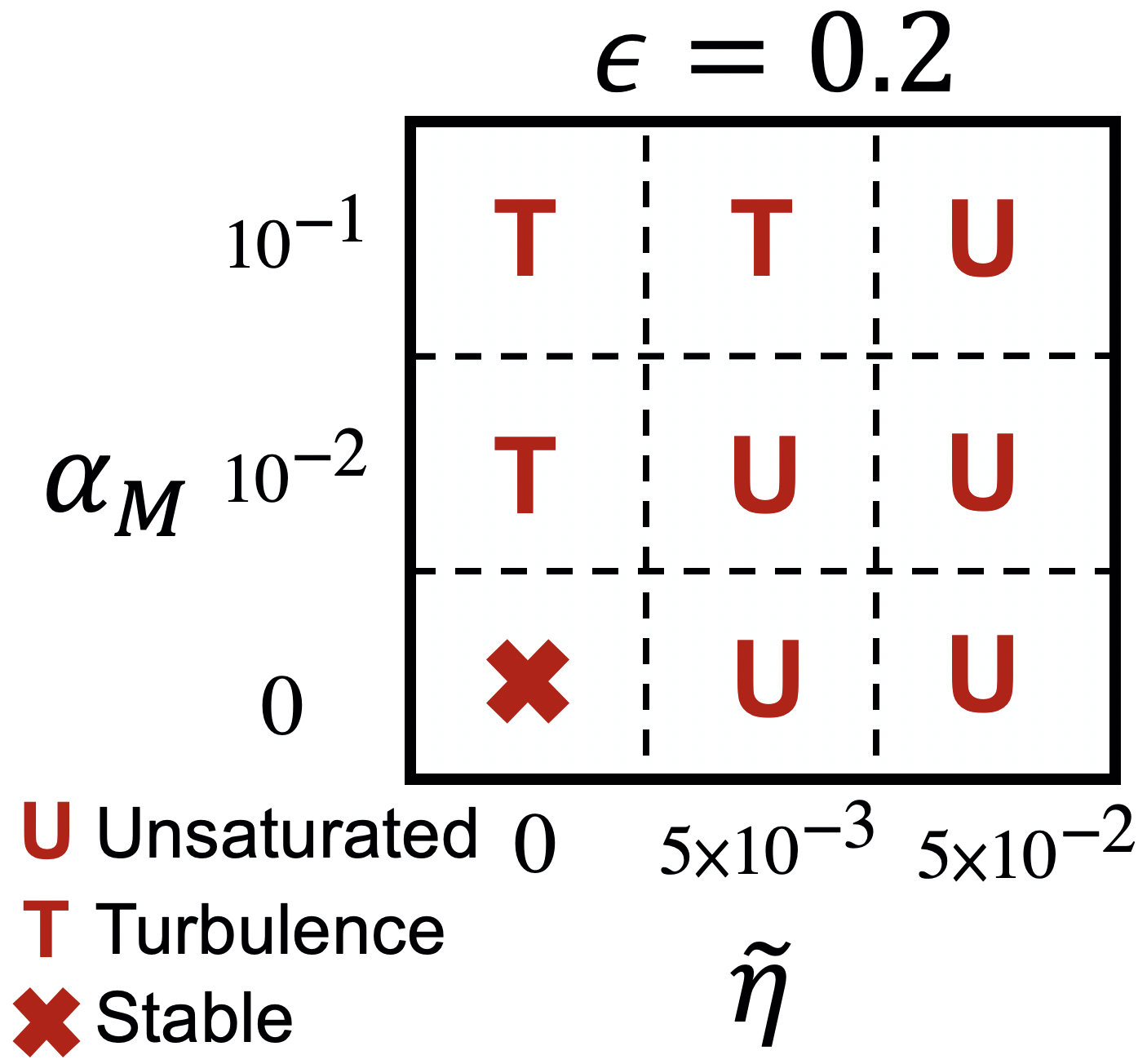}
    \caption{
   Similar to Fig. \ref{Paper_overview}, but for runs with $\epsilon=0.2$. 
    }
    \label{Paper_overview_02}
\end{figure}

Figs. \ref{Paper_overview}---\ref{Paper_overview_02} give a visual overview of the simulations listed in Table  \ref{table1} for different Maxwell stresses, $\alpha_M$, and global radial pressure gradients, $\etatilde$. We categorize our results based on the state at the end of the simulation. Clumping cases, denoted by `C', are those that saturate into a turbulent state with a maximum dust density exceeding the Roche density, i.e. Eq. \ref{strong_clump_cond} is met. Cases with `T' reach a turbulent, quasi-steady state but do not meet the clumping condition. 
Cases with `U' are unsaturated as the instability remains in its linear growth phase within the simulation timescale. Finally, cases marked with a cross ($\times$) are completely stable as the instability does not operate (namely when $\alpha_M =\etatilde=0$). 

\begin{figure*}
    \centering
    \includegraphics[width=\linewidth]{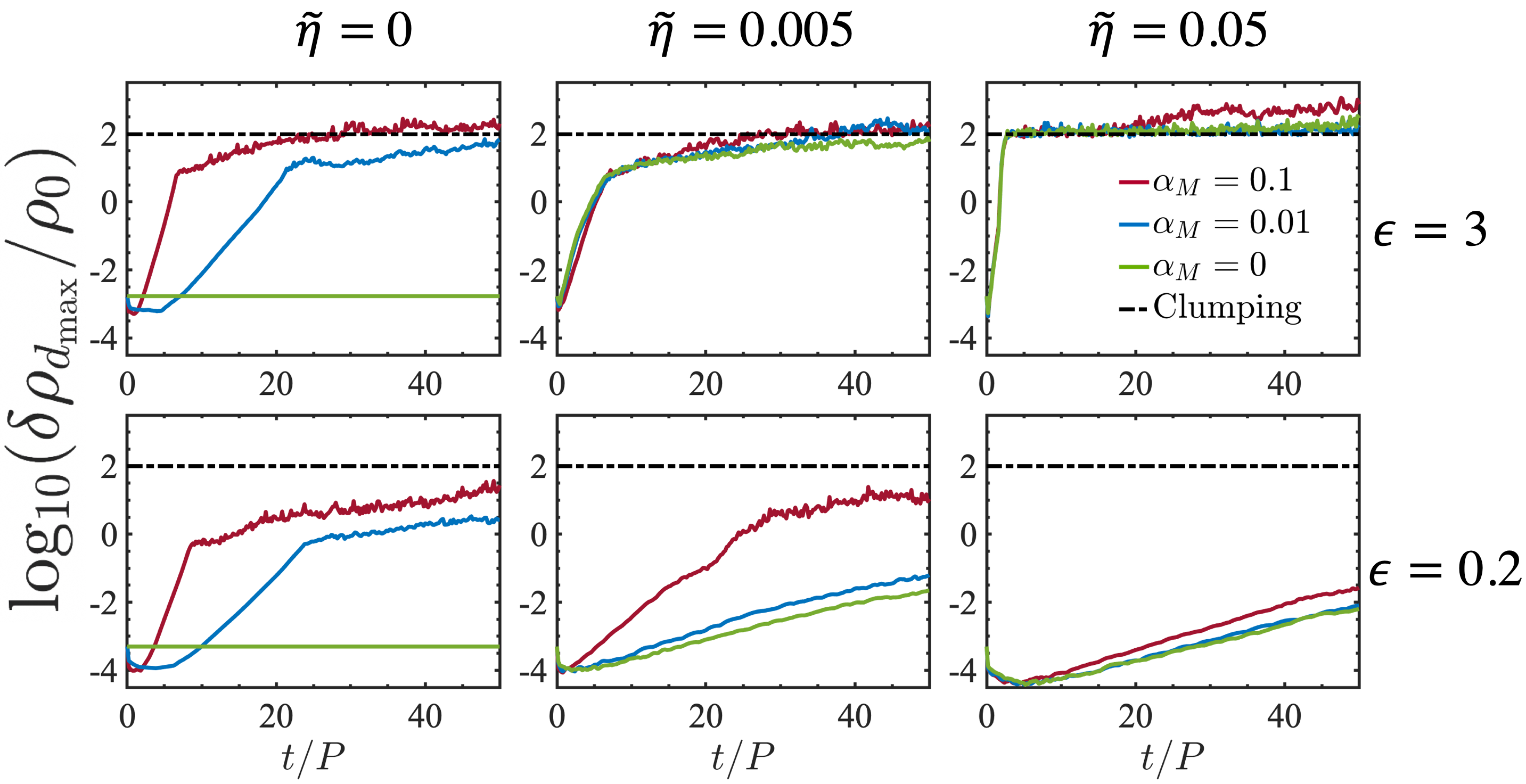}
    \caption{
    Time evolution of the maximum dust density perturbation for simulations listed in Table \ref{table1}. The upper and lower panels correspond to $\epsilon=3$ and $0.2$, respectively. The left, middle, and right columns corresponds to $\etatilde=0$ (no pressure gradient), $0.005$, and $0.05$, respectively. Within each panel we show vary the applied torque or Maxwell stress $\alpha_M=0$ (green, no accretion flow), $\alpha_M=0.01$ (blue), and $\alpha_M=0.1$ (red). The horizontal dashed-dotted line corresponds to the clumping condition, Eq. \ref{strong_clump_cond}.
    }
    \label{Paper_allgrowthrates}
\end{figure*}

Fig. \ref{Paper_allgrowthrates} shows the time evolution of the maximum dust density perturbation, $\delta\rhodmax\equiv\mathrm{max}\left(\rhod-\epsilon\rho_0\right)$, for the above simulations. The upper and (lower) panels show the $\epsilon = 3$ ($\epsilon=0.2$) cases. From left to right, the columns denote $\etatilde=0$, $0.005$, and $0.05$. The green, blue, and red curves denote $\alpha_M=0$, $0.01$, and $0.1$, respectively. We also mark the clumping condition (Eq. \ref{strong_clump_cond}) with the horizontal dashed-dotted line.

\subsection{Dust-rich disks}

Starting with $\epsilon=3$ and $\etatilde=0.05$, in all cases, $\rhodmax$ grows rapidly, with a growth rate $s \simeq 0.64\Omega$, and saturate into a turbulent state where the clumping condition is met. The case with $\alpha_M=0$ (no gas accretion) corresponds to the classic SI. We find the inclusion of a sufficiently strong accretion flow, here with $\alpha_M=0.1$, can further boost $\rhodmax$ by an order of magnitude. This enhancement is negligible for $\alpha_M=0.01$. 

Moving to weaker pressure gradients but still considering $\epsilon=3$, all cases become more stable as $\rhodmax$ is reduced. This is consistent with linear theory as in dust-rich disks growth rates drop with decreasing $\etatilde$ \citepalias{MK22}. { Since the classic SI (green curves) is powered by the radial pressure gradient, it is weakened with decreasing $\etatilde$ and} no longer meets the clumping condition with $\etatilde=0.005${,} and is stabilized altogether for $\etatilde=0$.  

{ The above result seemingly contradicts \cite{bai10c}'s finding that clumping via the SI is easier for decreasing (but non-zero) pressure gradients. However, a key difference is that their simulations are stratified. In that case, the weaker turbulence associated with a smaller pressure gradient allows particles to settle to a denser midplane layer, which ultimately promotes clumping. However, this settling effect is absent in unstratified simulations, so we observe SI-clumping for larger $\etatilde$.}

On the other hand, accreting disks ($\alpha_M>0$) are unstable for all $\etatilde$. For $\alpha_M=0.1$, the clumping condition is satisfied even if $\etatilde=0$, i.e. without a radial pressure gradient. { For $\epsilon=3$,} cases driven by the AdSI are insensitive to $\etatilde${, which differs from the dust-poor disks discussed in \S\ref{dust_poor}.}

\begin{figure*}
    \centering
    \includegraphics[width=\linewidth]{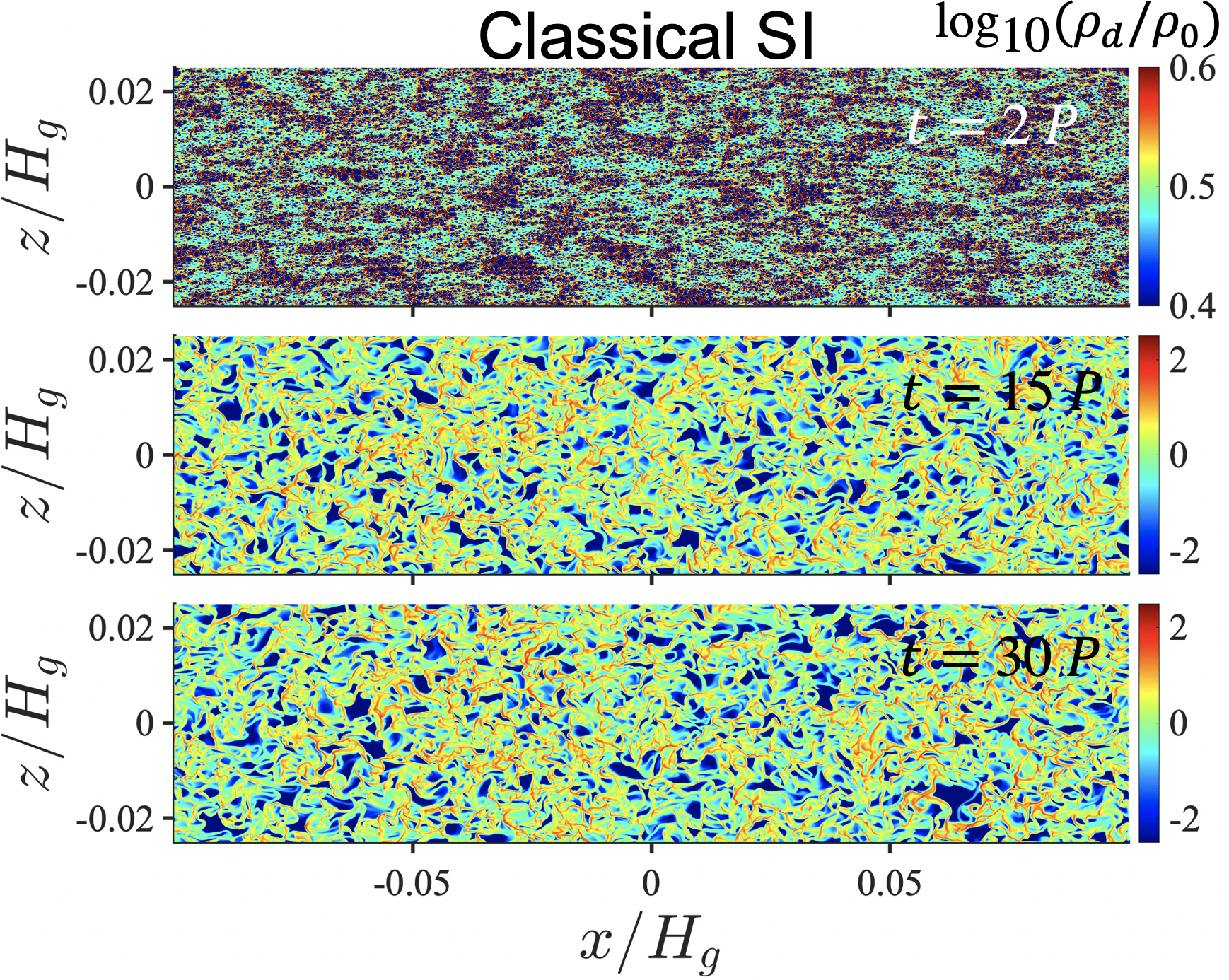}
    \caption{
    Dust density snapshots of the classic SI ($\epsilon=3$, $\etatilde= 0.05$, $\alpha_M=0$; run  E3eta005am0). The top panel shows the linear growth phase, while the middle and lower panels show the saturated state of the system. 
    }
    \label{Rhod3_in_space_005_0}
\end{figure*}

\begin{figure*}
    \centering
    \includegraphics[width=\linewidth]{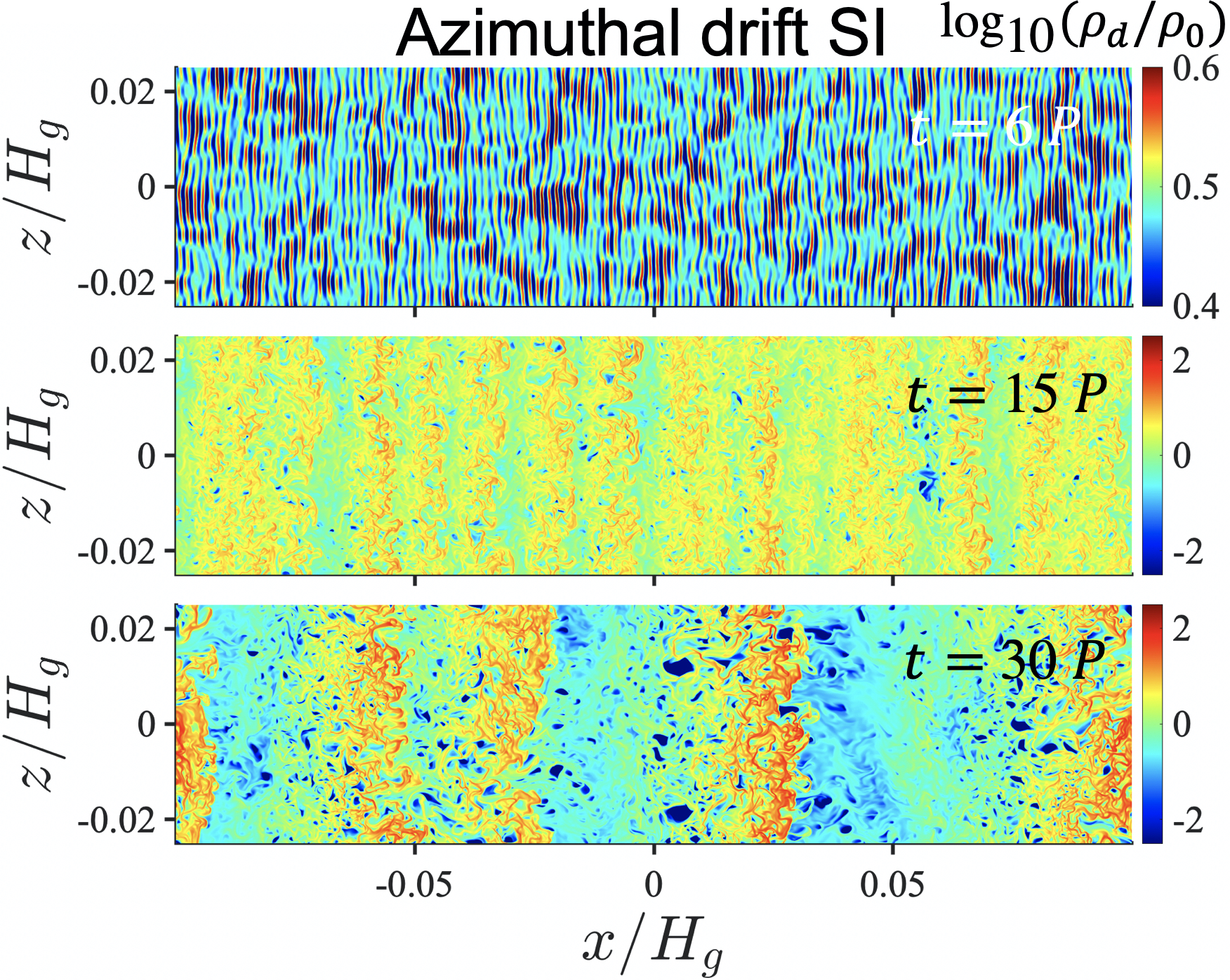}
    \caption{
    Dust density snapshots of the AdSI ($\epsilon=3$, $\etatilde=0$, $\alpha_M=0.1$, run E3eta0am01). 
    The top panel shows the linear growth phase, the middle panel shows the first saturated turbulent state, and the lower panel shows a second saturated state of the system with fewer vertical filaments (after merging) but with higher dust densities. 
    }
    \label{Rhod3_in_space_0_01}
\end{figure*}

Figs. \ref{Rhod3_in_space_005_0}---\ref{Rhod3_in_space_0_01} show dust density snapshots for runs E3eta005am0 (classical SI) and E3eta0am01 (AdSI), respectively, which display distinct evolution. The classic SI remains approximately isotropic from growth to saturation. Note that our small radial domains are not well-suited for capturing the long term evolution of classic SI filaments, which are typically separated by $0.2\Hgas$ (i.e. our box size) as found in large domain simulations \citep{yang14}. This is further discussed in \S\ref{discussion_filament}. 

By contrast, the AdSI shows anisotropy early on and is sustained. We find the preferential growth of vertically-extended filaments, initially with small radial separations. This is consistent with the linear theory developed by \citetalias{MK22} as the AdSI is intrinsically one-dimensional with little dependence on the vertical dimension. Such modes might then be expected to dominate numerical simulations as they should be more robust to grid dissipation than small-scale perturbations. 

In conjunction with Fig. \ref{Paper_allgrowthrates} (red curve in the top-left panel), we see that these vertical filaments grow by merging: at $t=15P$ the system reaches slowly-growing state with $\rhodmax\lesssim 10^2\rho_0$ and $\sim10$ filaments; while  by $t=30P$ as the system shifts into a second saturated phase  with $\rhodmax\gtrsim 10^2$ and we are left with $4$ filaments.

\subsection{Dust-poor disks}\label{dust_poor}

Next, we examine dust-poor disks with $\epsilon=0.2$, which are depicted in Fig. \ref{Paper_overview_02} and the bottom row of Fig. \ref{Paper_allgrowthrates}. (The dust density contour plots for this case are qualitatively similar to Fig. \ref{Rhod3_in_space_0_01}.) None of these runs meet the clumping condition within the simulation timescale. Furthermore, all of the classic SI cases ($\alpha_M=0$) remain in the linear growth phase. Nevertheless, we find that with $\alpha_M>0$, i.e. an accretion flow, dust 
can still be significantly concentrated if pressure gradients are weak. 

For $\etatilde=0.05$, all runs remain unsaturated. { Even in most unstable disk with $\alpha_M=0.1$, $\epsilon$ only increases by $\sim 1\%$}. Upon lowering to $\etatilde=0.005$, the $\alpha_M=0.1$ disk (red) saturate while the $\alpha_M=0.01$ disk (blue) still remains unsaturated. For $\etatilde=0$, both accreting disks reach saturation with $\rhodmax/\rho_0\sim O(1)-O(10)$, { which is} one to two orders of magnitude larger than the initial value.

The destabilization of the $\epsilon=0.2$ disks with decreasing $\etatilde$ is in direct contrast with the $\epsilon=3$ cases above, but is consistent with linear theory \citepalias{MK22}: in accreting, dust-poor disks, growth rates indeed increase with decreasing $\etatilde$ as the system transitions from the classic SI to the AdSI, but the opposite is true for dust-rich disks.

Our results demonstrate a qualitative difference between the AdSI and the classic SI at low dust-to-gas ratios, namely the AdSI can still drive dust concentrations by an order of magnitude or more. On the other hand, the classic SI, for example the `AA' run of \cite{johansen07}, with $(\epsilon, \st, \etatilde)=(0.2,0.1,0.05)$ only show about a $20\%$ increase in the maximum dust density perturbation, even after $300P$ in their runs. 


\subsection{Turbulence properties} \label{turbulence}


In this section, we investigate the kinetic energy spectra, angular momentum transport, and mass diffusion associated with SI-driven turbulence. Here, we are particularly interested in how the AdSI differs from the classical SI. To this end, we focus on the `C' and `T' cases listed in Table \ref{table1} where the system saturates into a turbulent state. Our diagnostics are described in \S\ref{Numerical_method} and Table \ref{table2} lists our measured values for the aforementioned runs.

\begin{table*}
\caption {Turbulence properties of simulations that reach saturation.
The columns from left to right are: name of the runs, radial pressure gradient, Maxwell stress, end state, the equilibrium and turbulent contributions to gas angular momentum transport, gas mass diffusion coefficients in each direction, and correlation times in each direction.
\label{table2}
} 
\begin{ruledtabular}
\begin{tabular}{llllllllllll}
  Name  & $\etatilde$ &  $\alpha_M$ & State & $\langle \overline{\alpha_{\rm SS}} \rangle_{\rm Eqm}  $ & $\langle \overline{\alpha_{\rm SS}} \rangle_{\rm Turb} $ &  $\alpha_{g,x}  $ &  $\alpha_{g,y} $ & $\alpha_{g,z} $ & $\tau_{c,x}$ & $\tau_{c,y} $ & $\tau_{c,z}$\\
        \hline
        \multicolumn{2}{c}{$\epsilon = 3$, $N_p = 2^{26}$} \\
        \hline
 E3eta005am0     & $0.05$  & $0$   & C & $         -2.35$e$-5$    & $        -7.61$e$-5 $ & $  4.36$e$-6 $ & $  2.87$e$-6 $ & $  2.21$e$-6 $ & $0.06$ & $0.09$ & $0.06$\\
 E3eta005am001    & $0.05$  & $0.01$& C & $        -2.19$e$-5 $    & $        -7.08$e$-5 $ & $  4.31$e$-6 $ & $  1.27$e$-6 $ & $  2.22$e$-6 $ & $0.06$ & $0.06$ & $0.06$\\
 E3eta005am01     & $0.05$  & $0.1$ & C & $        -7.81$e$-6 $    & $\; \; \, 7.65$e$-5 $ & $  7.04$e$-6 $ & $  1.13$e$-3 $ & $  4.37$e$-4 $ & $0.06$ & $1.47$ & $1.02$\\
        \cline{2-4}
 E3eta0005am0     & $0.005$ & $0$   & T & $        -2.35$e$-7 $    & $        -3.58$e$-7 $ & $  3.36$e$-8 $ & $  1.32$e$-7 $ & $  2.87$e$-7 $ & $0.09$ & $0.21$ & $0.36$\\
 E3eta0005am001   & $0.005$ & $0.01$& C & $        -7.81$e$-8 $    & $\; \; \, 2.36$e$-7 $ & $  1.42$e$-7 $ & $  6.56$e$-6 $ & $  2.04$e$-7 $ & $0.18$ & $1.14$ & $0.24$\\
 E3eta0005am01    & $0.005$ & $0.1$ & C & $\; \; \, 1.38$e$-6 $    & $\; \; \, 8.40$e$-6 $ & $  4.77$e$-7 $ & $  5.58$e$-5 $ & $  6.79$e$-7 $ & $0.09$ & $1.23$ & $0.15$\\
        \cline{2-4}

 E3eta0am001      & $0$     & $0.01$& T & $\; \; \, 5.87$e$-10$    & $\; \; \, 3.27$e$-8 $ & $  7.27$e$-9 $ & $  2.44$e$-7 $ & $  4.96$e$-8 $ & $0.21$ & $1.11$ & $0.51$\\
 E3eta0am01     & $0$     & $0.1$ & C & $\; \; \, 5.87$e$-8 $    & $\; \; \, 7.73$e$-6 $ & $  5.37$e$-7 $ & $  5.62$e$-5 $ & $  1.32$e$-6 $ & $0.09$ & $1.32$ & $0.21$\\
       \hline
        \multicolumn{2}{c}{$\epsilon = 0.2$, $N_p = 2^{22}$} \\
        \hline

 E02eta0005am01   & $0.005$ & $0.1$ & T & $\; \; \, 1.70$e$-5 $    & $\; \; \, 4.45$e$-6 $ & $  1.40$e$-6 $ & $  2.84$e$-5 $ & $  1.61$e$-6 $ & $0.15$ & $0.72$ & $0.12$\\
        \cline{2-4}
 E02eta0am001     & $0$     & $0.01$& T & $\; \; \, 1.44$e$-9 $    & $\; \; \, 1.24$e$-8 $ & $  4.86$e$-9 $ & $  1.73$e$-7 $ & $  2.75$e$-8 $ & $0.18$ & $0.84$ & $0.27$\\
 E02eta0am01.     & $0$     & $0.1$ & T & $\; \; \, 1.44$e$-7 $    & $\; \; \, 2.08$e$-6 $ & $  3.89$e$-7 $ & $  4.65$e$-5 $ & $  4.39$e$-7 $ & $0.09$ & $1.47$ & $0.09$\\

\end{tabular}
\end{ruledtabular}
\end{table*}

\subsubsection{Kinetic energy spectra}

\begin{figure}
    \centering
    \includegraphics[width=\linewidth]{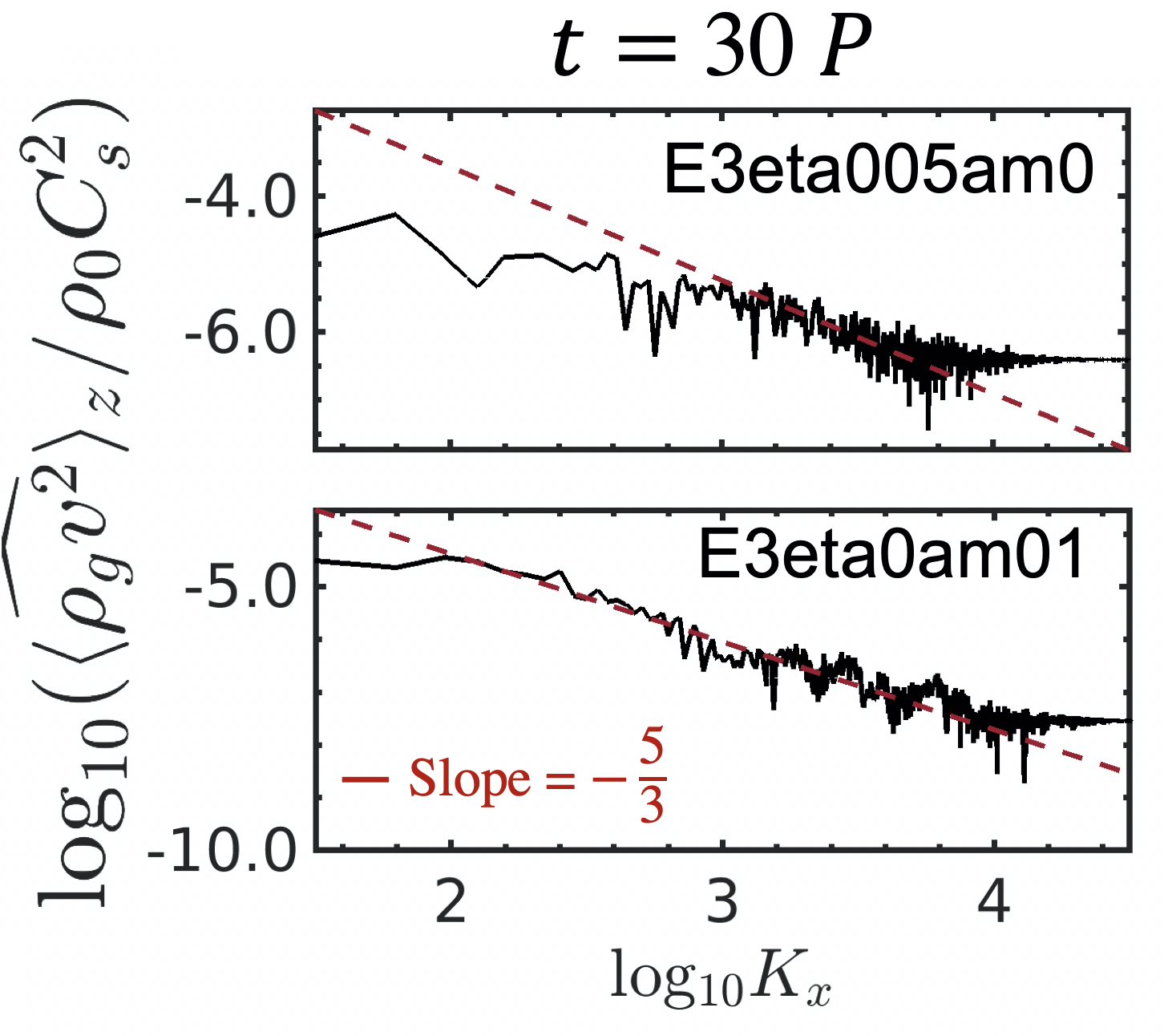}
    \caption{
    Spectra of the vertically-averaged gas kinetic energy density for the classic SI (top, run E3eta005am0) and the AdSI (bottom, run E3eta0am01). The red dashed lines mark a slope of -5/3. 
    }
    \label{Kolmo}
\end{figure}

Fig. \ref{Kolmo} compares the gas kinetic energy spectrum for runs E3eta005am0 (classic SI) and E3eta0am01 (AdSI) on a logarithmic scale at $t=30P$, when both systems are in quasi-steady state (see Fig. \ref{Paper_allgrowthrates}). The red dashed lines denote a slope of $-5/3$, i.e. the Kolmogorov law \citep{kolmogorov41}. We find the AdSI follows the Kolmogorov spectrum from $K_x\sim 10^2$--$10^4$ and is hence the inertial range; but the classic SI only from $K_x\sim 10^3$--$10^4$. Note that $K_x=10^4$ corresponds to a wavelength of $6\times10^{-4}H_g$, which is resolved by about $6$ cells. Larger wavenumbers are thus not well-resolved and the associated dynamics cannot be properly captured. This explains the deviation from the Kolmogorov spectrum at small scales. 


The above spectra are consistent with the contour plots shown in Fig. \ref{Rhod3_in_space_005_0} and \ref{Rhod3_in_space_0_01}. Namely, the former classic SI case show small-scale turbulence, while the latter AdSI case shows large-scale vertical filaments that dominate the system, as well as small-scale eddies within them. 

However, we caution that the above result for the classic SI may be affected by the domain size. According to \cite{yang14}, classic SI filaments have radial separations of order $0.2H_g$ (our box size). Thus, increasing the domain is expected to support larger scales, and possibly extend the match with the Kolmogorov law to smaller $K_x$. 

\subsubsection{Angular momentum transport}

We quantify the radial flux of gas orbital momentum with $\alpha_{\rm SS}$, as described in \S\ref{Diagnostics}, where a positive value indicates outwards transport. These are listed in the 5th and 6th columns of Table \ref{table2}, where we further decompose $\alphass$ into that associated with the initial equilibrium and deviations from it (or the turbulent part). The equilibrium value is calculated from Eq. \ref{gas_stress} using Eqs. \ref{eqm_vr} and \ref{eqm_vphi}; while the perturbed part is obtained from Eq. \ref{gas_stress_time_aver} and subtracting the  equilibrium part. 

First, we point out that the equilibrium transport is negative when dust-gas drift is dominated by the radial pressure gradient \citep[as noted by][]{johansen07}; while for sufficiently large $\alpha_M/\etatilde$, the background transport becomes positive. 

We find that in most cases if the equilibrium $\alphass$ is negative, the perturbed part is also negative, indicating inwards transport by the classic SI. However, a sufficiently strong accretion flow can reverse the direction of angular momentum transport, as observed for runs E3eta0005am001 and E3eta005am01. In these cases, the total transport is positive, although the background is negative. For cases with a positive background transport, i.e. when the azimuthal drift becomes dominant, the perturbed transport is also positive. We conclude that the AdSI drives outwards angular momentum transport in the gas. 

Consider now the $\epsilon=3$ cases. As above, at fixed $\etatilde$, transport becomes positive and increases in magnitude as $\alpha_M$ increases. Similarly, at a given $\alpha_M$ the magnitude of transport increases with $\etatilde$. 
For the classic SI ($\alpha_M=0$), the equilibrium and turbulent parts of $\alphass$ have comparable magnitudes, though the latter is larger. However, with increasing $\alpha_M$ at fixed $\etatilde$, the turbulent contribution to $\alphass$ well-dominates the transport. For example, for E3eta0am01 the turbulent-to-equilibrium transport ratio is $O(10^2)$. However, the total transport is still relatively weak with $\alphass\lesssim 10^{-5}$.

For dust-poor disks with $\epsilon=0.2$ and $\etatilde=0$ (so the system is driven by AdSI), we still find the turbulent transport dominates the equilibrium value, but here only by a factor $\sim 10$. Curiously, we find the run E02eta0005am01, with a weak pressure gradient of $\etatilde=0.005$, the turbulent transport is sub-dominant. This suggests that the classic SI may have non-negligible (negative) contributions in this case. 


\subsubsection{Mass diffusion}

In Table \ref{table2}, we calculate the bulk diffusion coefficients for the gas in the $i^\mathrm{th}$ direction, $\alpha_{g,i}$, and list the corresponding dimensionless correlation times, $\tau_{c,i}$. These are related via the velocity dispersion $\delta v_i$, see Eq. \ref{gas_diffusion_coef}. For tightly-coupled dust  with $\st\ll 1$, we expect gas and particle diffusion coefficients to be equivalent \citep{youdin07b,youdin11}.

For $\etatilde=0.05$ and $\alpha_M\leq0.01$, we find $\alpha_{g,i}$ of $O(10^{-6})$ and is approximately isotropic. However, the strongly torqued disk with $\alpha_M=0.1$ (E3eta005am01) is clearly anisotropic and is dominated by $\alpha_{g,y}$ of $O(10^{-3})$. Such an anisotropy with an enhanced azimuthal diffusion is exemplified in torqued disks with weak (including zero) pressure gradients. For example, in the pure AdSI run E3eta0am01, we find 
$\alpha_{g,y}$ of $O(10^{-5})$, which is two orders of magnitude larger than $\alpha_{g,x}$. This indicates that while the classic SI turbulence is approximately isotropic, AdSI turbulence is anisotropic. 

\begin{figure*}
    \centering
    \includegraphics[width=\linewidth]{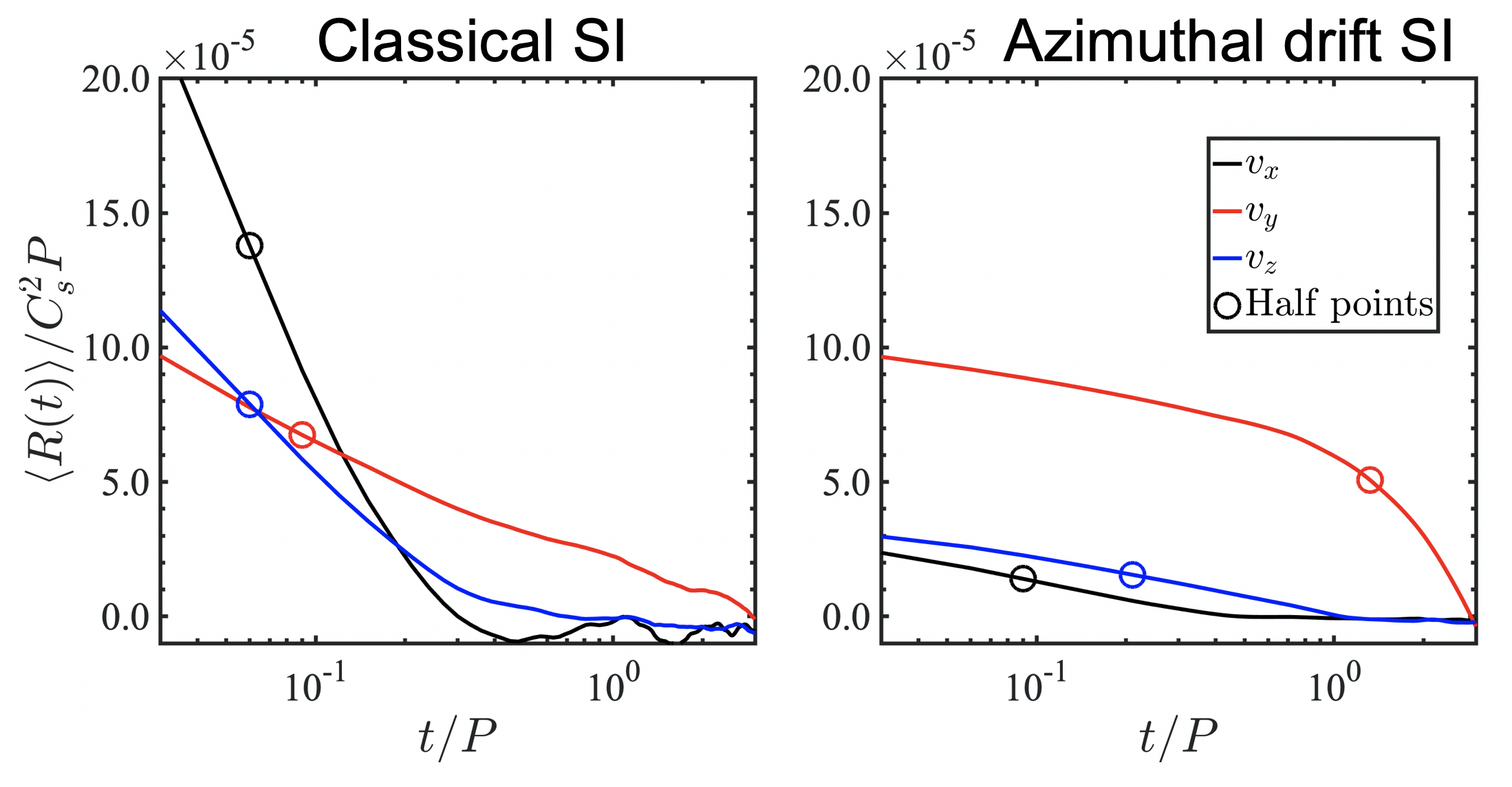}
    \caption{{ 
    Auto-correlation function of the gas velocity fluctuations for runs the classic SI (left, run E3eta005am0) and the AdSI (right, run E3eta0am01).
    Black, red, and blue lines denote the $x$, $y$, and $z$ velocity components, respectively. 
    Circles are the half-life points for each component of the velocity fluctuations, which defines the correlation times $t_{c,i}$.} 
    }
    \label{AutoC_plot}
\end{figure*}

We find that longer correlation times in $v_y$ is the dominant cause of anisotropy in torqued disks. Fig.\ref{AutoC_plot} shows the auto-correlation function of the gas velocity fluctuations for runs E3eta005am0 (classic SI) and E3eta0am01 (AdSI). In the latter case, the profile of the $v_y$ auto-correlation function differs significantly from that for $v_x$ and $v_z$. In the plots, circles denote the half-life of the auto-correlation functions. From these we obtain correlation times $\tau_{c,i} = 0.06P, 0.09P, 0.06P$ in the $x,\,y,\,z$ velocities, respectively, for the classic SI; and $\tau_{c,i}=0.09P,\, 1.32P,\, 0.21P$ for the AdSI. 

{ Overall, AdSI correlation times are longer, especially in the azimuthal and vertical velocities, which are larger by a factor of $\sim 10$ and $\sim 3$ than the classic SI, respectively. However, while the corresponding $\alpha_{g,y}$ for the AdSI is also larger by about an order of magnitude; $\alpha_{g,z}$ is slightly smaller, and $\alpha_{g,x}$ is significantly smaller than the classic SI, see Table \ref{table2}. This suggest weaker turbulent stirring in the $(x,z)$ plane with smaller meridional velocity fluctuations (Eq.  \ref{gas_diffusion_coef}).}

\subsection{Radial drift of dust}

\begin{figure*}
    \centering
    \includegraphics[width=\linewidth]{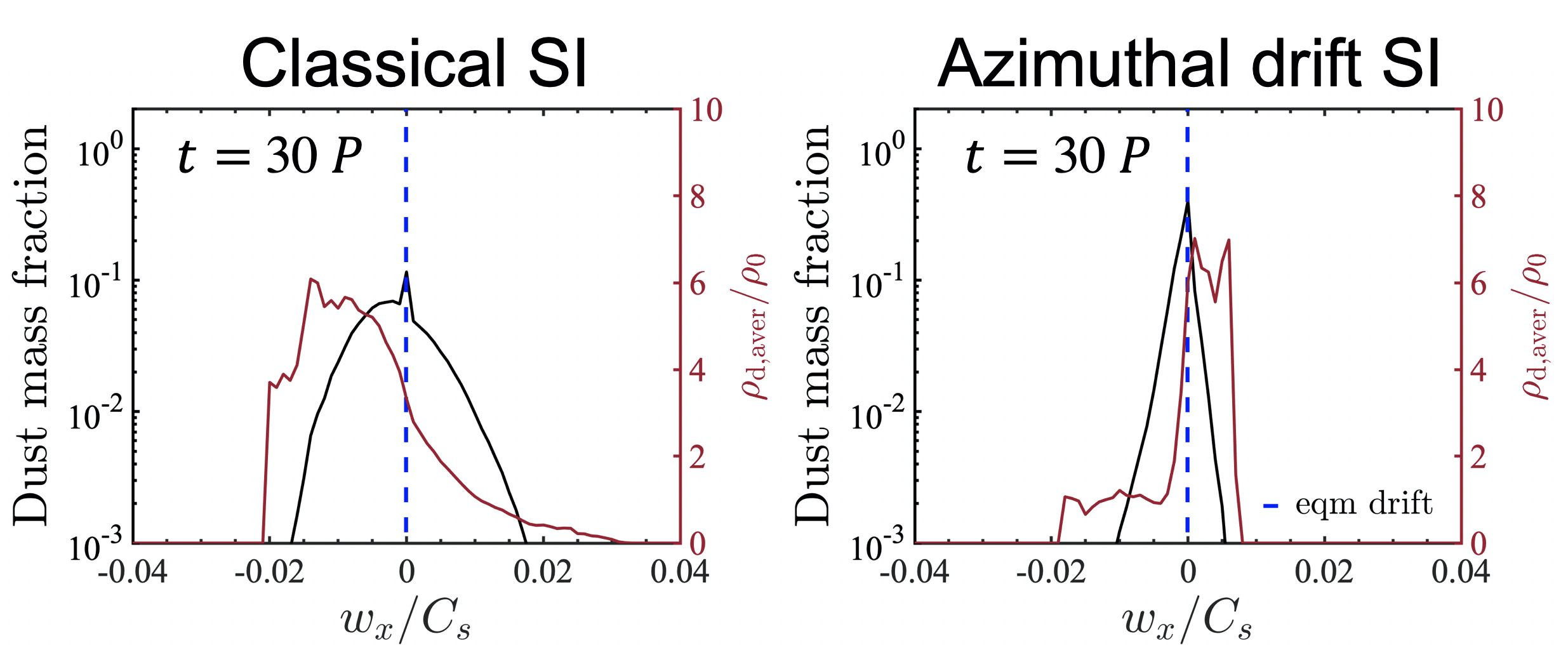}
    \caption{
    { The fraction of the total dust mass with a given radial velocity $w_x$ (black, left axis) and the average dust density associated with a given $w_x$ (red, right axis), for the classic SI  (left panel) and the AdSI (right panel). The vertical, blue dashed line corresponds to the equilibrium drift velocities in each case.}
    }
    \label{transport_wx}
\end{figure*}

We compare the drift of solids between the classic SI (E3eta005am0) and the AdSI (E3eta0am01) { when the systems are in a quasi-steady turbulent state at $t=30P$.} We follow a similar methodology as \cite{johansen07}, { who} models dust as Lagrangian particles and counts the number of particles with a given velocity and their average ambient density. { Here, we sum the dust mass from grid cells with $w_x$ to $w_x+10^{-3}C_s$, then divide by the total dust mass to obtain the mass fraction of dust in a given radial velocity bin. We also calculate the average dust density in each bin. The result is shown in Fig. \ref{transport_wx}. 
}

For the classic SI, we obtain similar results as  \citeauthor{johansen07} for tightly coupled grains. Namely, the distribution is approximately Gaussian with high dust { densities} picking up larger inwards drift { speeds}. This is opposite to the equilibrium drift solution (Eq. \ref{eqm_wr} with $\alpha_M=0$), which predicts slower drift with increasing dust-to-gas ratio. { This can be explained by high-density dust clumps 
experiencing a weaker gas drag as it is only subject to drag on their surface, while grains inside the clump are shielded from the exterior gas. This results in a dust clump having an effectively longer stopping time than an individual dust grain \citep{johansen07}. 
} 

By contrast, the distribution for the  AdSI is somewhat negatively skewed with low dust { densities} having the fastest inwards drift. This is in fact consistent with the equilibrium drift given by Eq. \ref{eqm_wr} (with $\etatilde=0$), since for $\st\ll 1$,
\begin{align}
w_x\propto -\frac{\alpha_M\hgas}{1+\epsilon}C_s \quad \text{($\etatilde=0,\, \st\ll1$)}.  \label{adsi_drift}
\end{align}
Thus $w_x$ becomes more negative with decreasing $\epsilon$. Here, dust is dragged inwards by the accreting gas. Notice the above expression is independent of $\st$. Thus, an increased effective stopping time for a dust clump does not affect its drift speed. 
Instead, the increased $\epsilon$ should slow down drift. Indeed, higher dust density regions have smaller $|w_x|$, but regions with $w_x>0$ cannot be explained with the equilibrium drift solution above. 

The AdSI result shares some resemblance with cases of the classic SI for $\st=1$ as considered by \citeauthor{johansen07}. As noted by \cite{youdin07}, for marginally coupled grains, the azimuthal drift also becomes non-negligible even for the classic SI. This suggests that azimuthal drift makes a key difference in the behavior of dust clumps in the turbulent state. 


\subsection{Dust concentrations} \label{Dust_concentrations}


We examine the propensity for the dust to concentrate or clump in SI-turbulent disks.  Recall that our models do not include self-gravity and we assess whether or not  gravitational collapse would occur based on the Roche density given by Eq. \ref{strong_clump_cond}. 

All of the clumping cases in our dust-rich ($\epsilon=3$) runs are in the upper right region of Fig. \ref{Paper_overview}, which indicates that a sufficiently large $\etatilde$, $\alpha_M$, or both, can concentrate dust efficiently.

None of our dust-poor ($\epsilon=0.2$) runs achieve strong clumping. However, we find that dust can still concentrate significantly if $\alpha_M/\etatilde$ is sufficiently large. These correspond to the turbulent cases shown in the upper left region of Fig. \ref{Paper_overview_02}. For reference, the run with $(\epsilon, \etatilde, \alpha_M)=(0.2, 0, 0.1)$ attains $\rho_\text{d, max} \simeq 44\rho_0$ in the saturated state, which is $\sim 200$ times larger than the initial dust-to-gas ratio. For $(\epsilon, \etatilde, \alpha_M)=(0.2, 0, 0.01)$ we find $\rho_\text{d, max} \simeq 3\rho_0$, which is still an order-of-magnitude enhancement. However, increasing $\etatilde\to0.005$ for this case results in an unsaturated state. Thus, in dust-poor accreting disks, the radial pressure gradient works against dust concentrations.

\begin{figure}
    \centering
    \includegraphics[width=\linewidth]{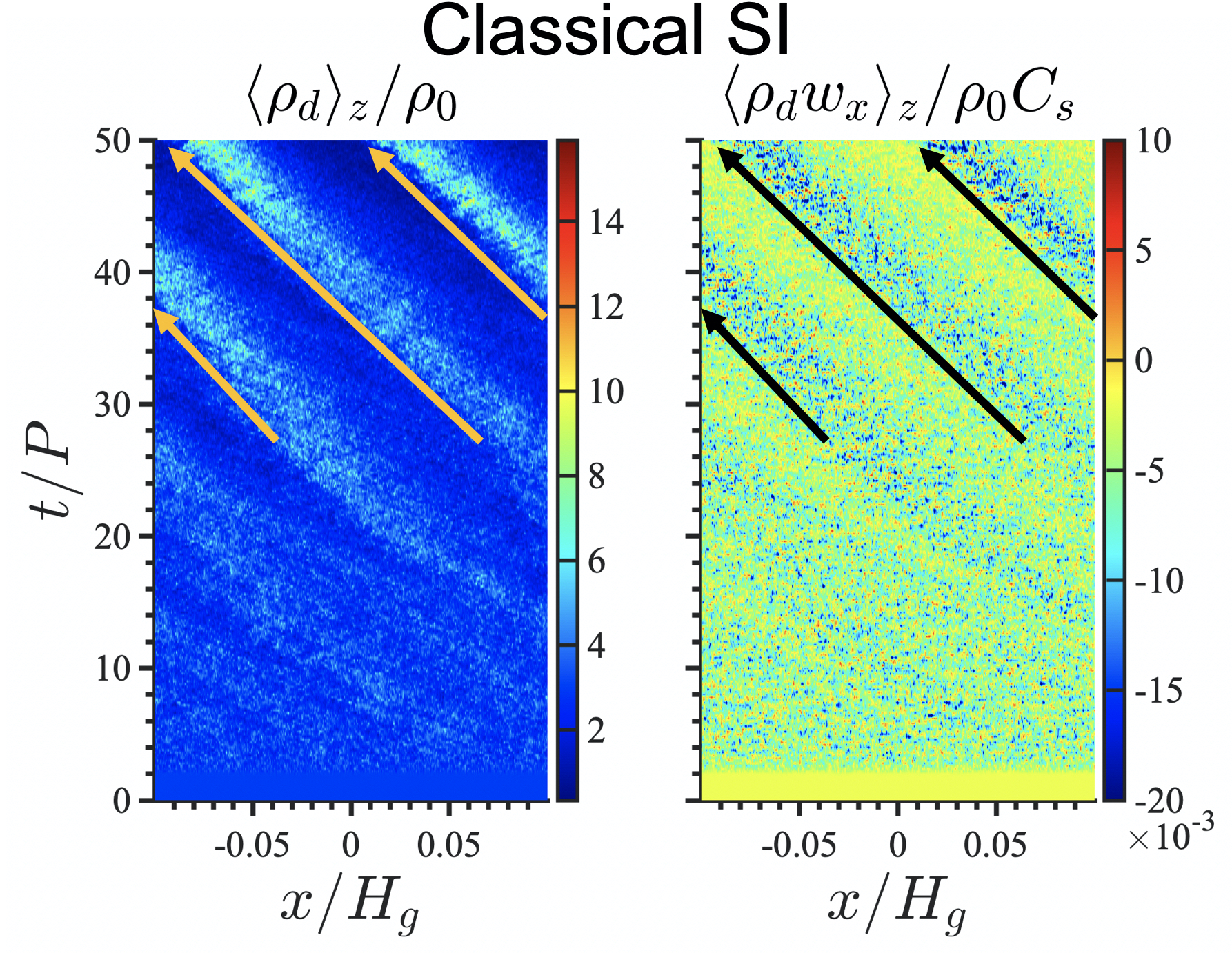}
    \caption{
    Time evolution of the vertically-averaged dust density (left) and radial dust mass flux (right) for the classic SI (run E3eta005am0). Arrows are drawn by inspection and indicate the drift of dust filaments. 
    }
    \label{transport_classical}
\end{figure}

\begin{figure}
    \centering
    \includegraphics[width=\linewidth]{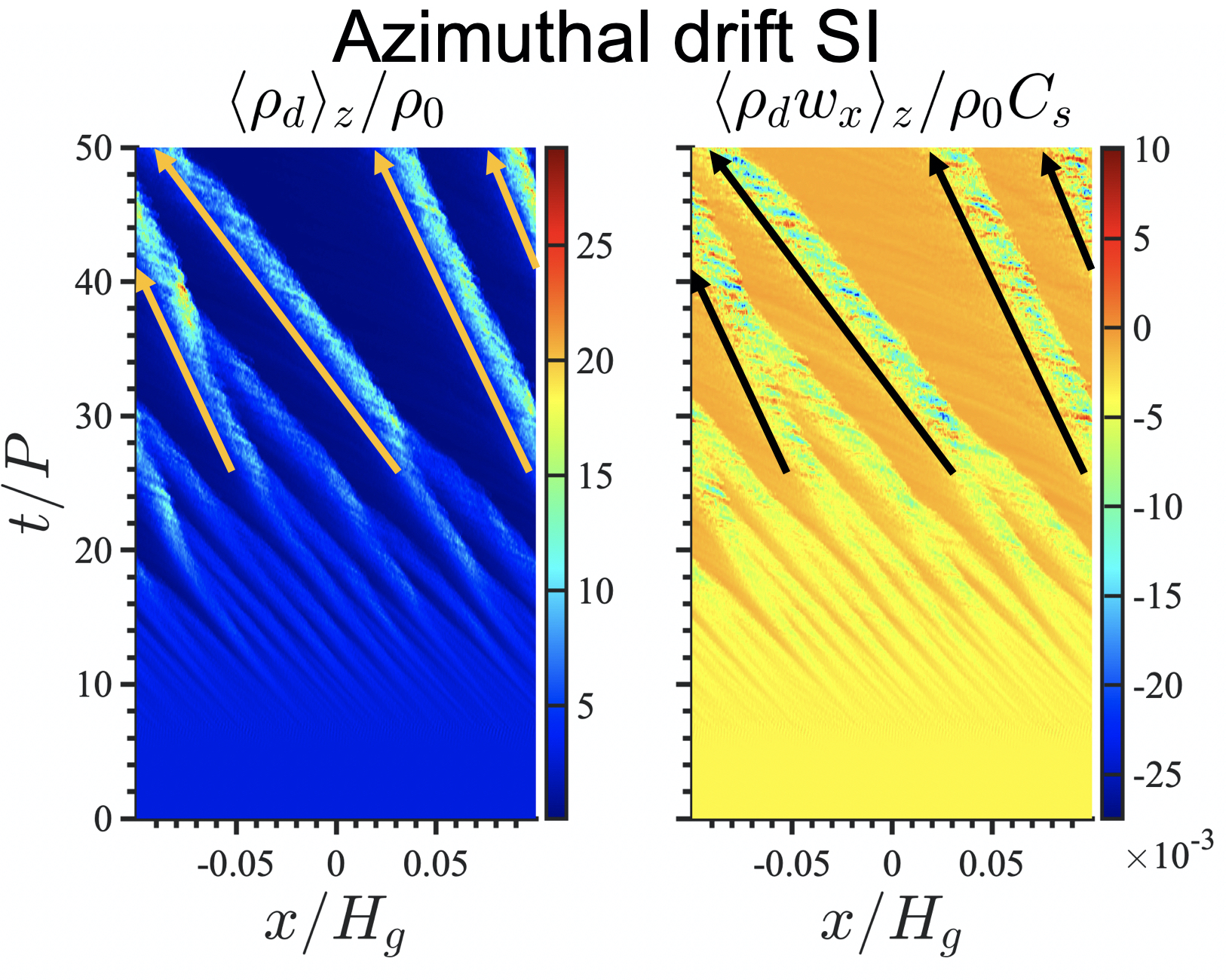}
    \caption{
    Same as Fig. \ref{transport_classical}, but for the AdSI (run E3eta0am01). 
    }
    \label{transport_azimuthal}
\end{figure}

Figs. \ref{transport_classical}---\ref{transport_azimuthal} shows the time evolution of the vertically-averaged dust density and radial dust mass flux for runs E3eta005am0 (classical SI) and E3eta0am01 (AdSI), respectively. These space-time plots show the formation and evolution of dust filaments. Black arrows are drawn (by inspection) to indicate their movement. These large-scale filaments are not easily discernible in snapshots of the classic SI (Fig. \ref{Rhod3_in_space_005_0}), but appear after vertically averaging the fields. 

For the classic SI, Fig. \ref{transport_classical} shows the emergence of two dust filaments at $t\simeq 30P$ with a separation of $0.1\Hgas$. They have azimuthal velocities closer to Keplerian values (which corresponds to $w_y=0$) than the ambient disk. We find the filaments slowly move inwards with a velocity of $w_{x, {\rm gp}} \sim -9 \times 10^{-4}C_s$. Note that this is the filament's group velocity (i.e. the gradient of the arrows) and not the dust fluid's radial velocity within the clump. These two filaments do not merge within the run. 

By contrast, in the AdSI case, Fig. \ref{transport_azimuthal} shows that multiple filaments emerge at the beginning of the nonlinear state ($t\simeq 10P$). These small-scale filaments then undergo pair-wise merging and eventually the system is left with three filaments. This process is also visible in the middle and bottom panels of Fig. \ref{Rhod3_in_space_0_01}.  Each merging event also increases the local dust density, which again differs from the classic SI case where each filament becomes denser individually. Notice there that the filaments' group radial velocity is still negative, despite regions of outwards dust motions within them.

\begin{figure}
    \centering
    \includegraphics[width=\linewidth]{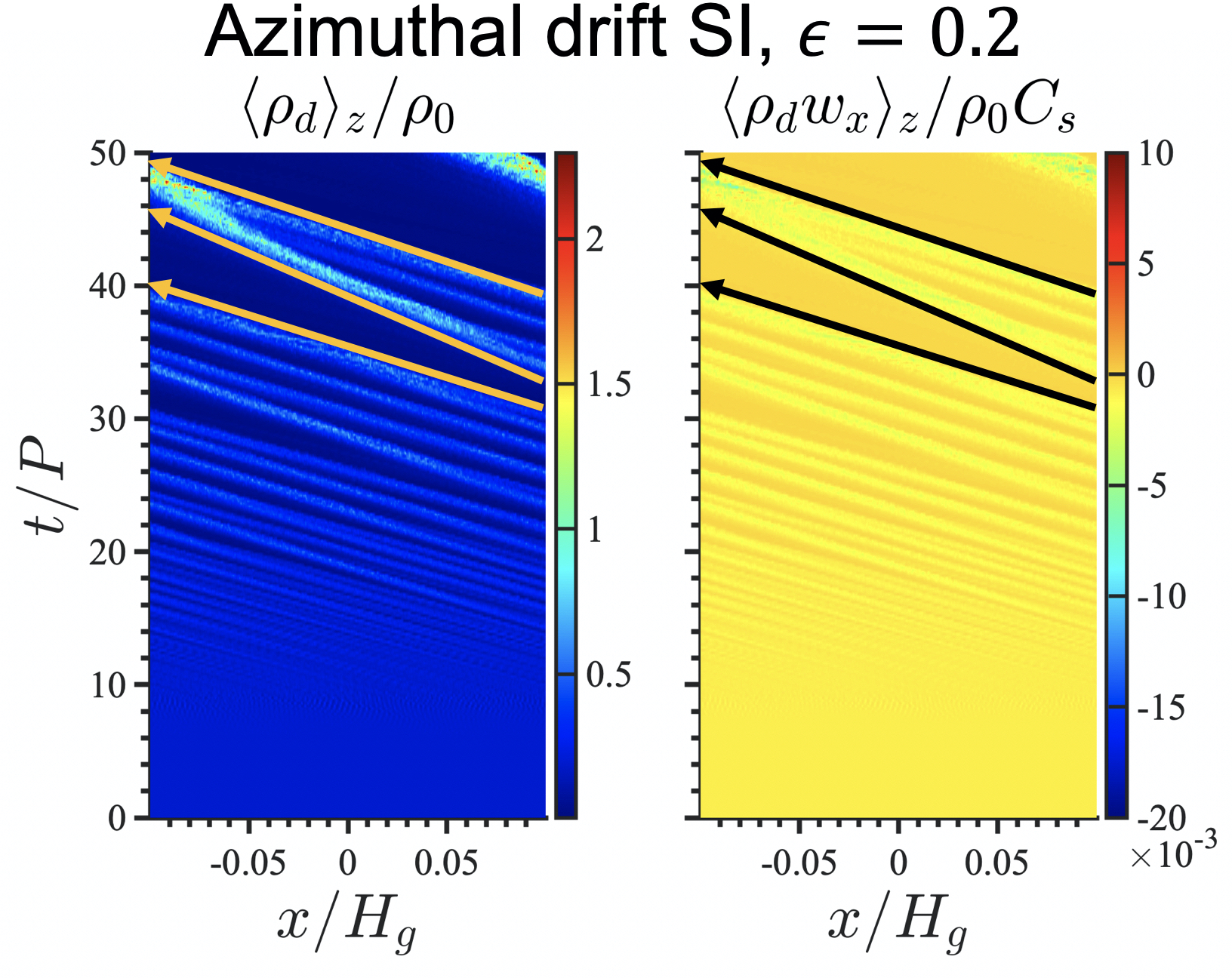}
    \caption{
    Same as Fig. \ref{transport_azimuthal}, but for the AdSI in dust-poor disk with $\epsilon=0.2$ (run E02eta0am01). 
    }
    \label{transport_azimuthal_eps02}
\end{figure}

For completeness, we show in Fig.  \ref{transport_azimuthal_eps02} the space-time plot of the vertically-averaged dust density and radial mass flux for an AdSI run with $\epsilon=0.2$ ( E02eta0am01). As in the dust-rich case, multiple filaments emerge from the linear instability, but now with much faster inwards drift speeds, which is consistent with the torque-induced drift given by Eq. \ref{adsi_drift}, which increases in magnitude with decreasing $\epsilon$. As seen in the figure with the middle yellow arrow, this reduction facilitates merger events as a denser filament's drift is reduced, it can capture incoming, lighter filaments with faster inward drifts. 



\subsection{AdSI in viscous disks}\label{addsims}



In the limit of vanishing gas viscosity and dust diffusion, both the classic SI and the AdSI can grow on arbitrarily small scales \citepalias{MK22}. 
This means that in inviscid simulations the system is always unstable on the grid scale. Here, we re-run several simulations with a small viscosity $\nu = \alpha_\mathrm{visc}C_s\Hg$ and a diffusion coefficient $D$ of the same value, to check that our main results on the AdSI are unaffected by under-resolved modes at the grid scale. 
These simulations were extended slightly to capture filament merging, which was found to affect the maximum dust densities.

\begin{figure}
    \centering
    \includegraphics[width=\linewidth]{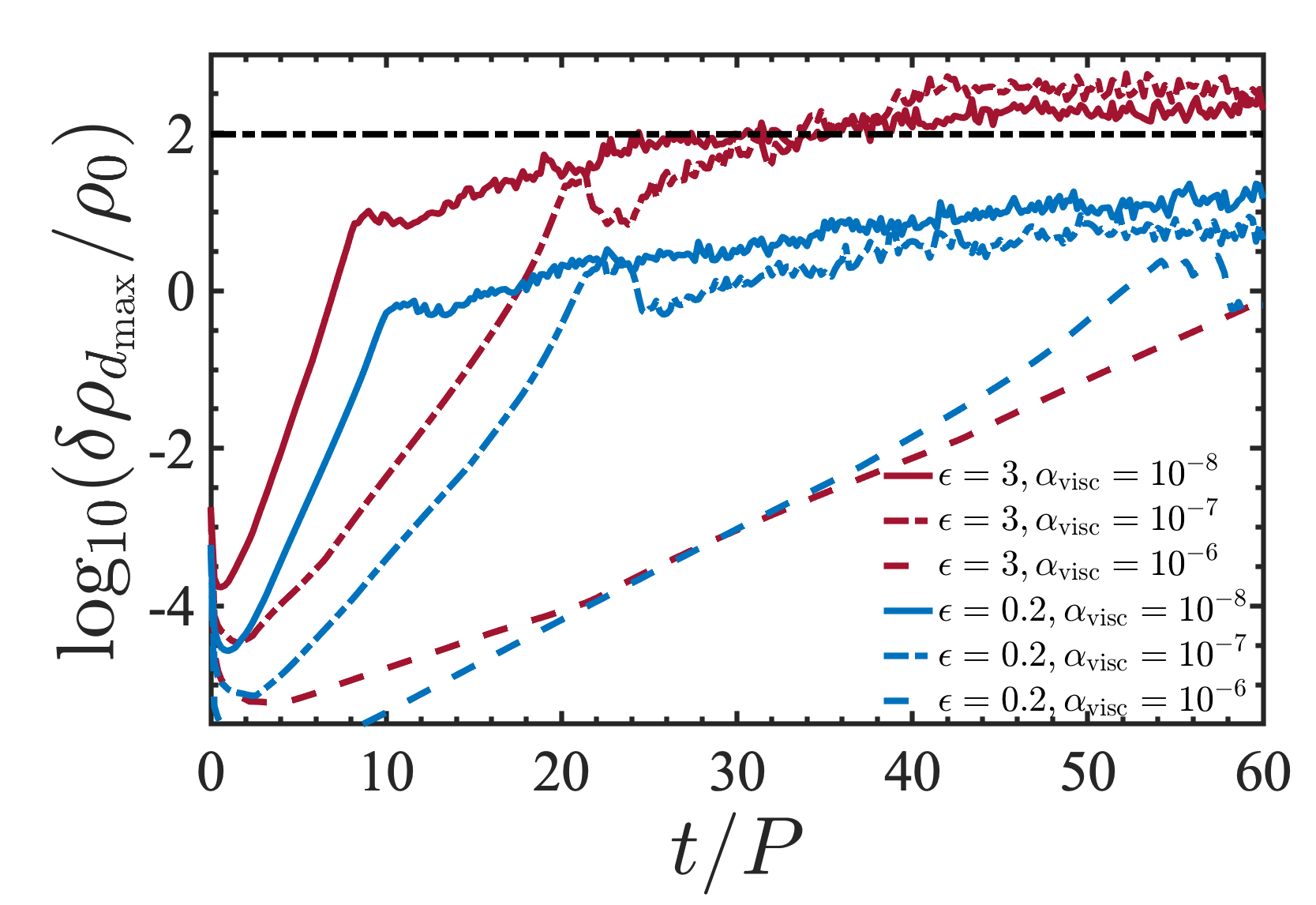}
    \caption{
    Time evolution of the maximum dust density perturbation in viscous simulations of the AdSI with $\etatilde=0$ and $\alpha_M=0.1$. The red lines and blue lines  corresponds to $\epsilon=3$ and $0.2$, respectively. The solid lines, { dashed-dotted lines, and dashed lines }corresponds to $\alpha_\mathrm{visc}=10^{-8}$, { $\alpha_\mathrm{visc}=10^{-7}$, and $\alpha_\mathrm{visc}=10^{-6}$}, respectively. 
    The horizontal dashed-dotted line corresponds to the clumping condition, Eq. \ref{strong_clump_cond}.
    }
    \label{Paper_growth_vis}
\end{figure}

Fig. \ref{Paper_growth_vis} shows the time evolution of the maximum dust density perturbation for selected runs with $\etatilde=0$, $\alpha_M=0.1$ with $\alpha_{\rm visc}=10^{-8}$ (solid), { $\alpha_{\rm visc}=10^{-7}$ (dashed-dotted), and $\alpha_{\rm visc}=10^{-6}$ (dashed)}. The red and blue curves correspond to $\epsilon=3$ and $0.2$, respectively. 

Unsurprisingly, a larger viscosity prolongs the linear phase of the instability, as growth rates are reduced and small-scale modes are suppressed (\citetalias{MK22}, see also Appendix \ref{code_test}). { In fact, for $\alpha_\mathrm{visc}=10^{-6}$, the $\epsilon=0.2$ disk only just saturates at $\delta\rho_\mathrm{d,max}/\rho_0\sim 1$ and the $\epsilon=3$ disk is still in its linear growth phase at the end of the simulations\footnote{{ We were not able to further extend the $\alpha_\mathrm{visc}=10^{-6}$ simulations due to the computational cost.}}. In the discussion below, we focus on $\alpha_\mathrm{visc}=10^{-8}$ and $10^{-7}$, for which the AdSI grows and saturates into a quasi-steady turbulent state. 
}


Consider the $\epsilon=3$ runs. We find for both { $\alpha_\mathrm{visc}=10^{-8}$ and $10^{-7}$} the system meets the clumping condition with $\rho_\text{d,max} \sim 550\rho_0$ at the end of the runs. 
However, there is a phase ($t=40$---$58P$) in which the higher viscosity run with $\alpha_\text{visc}=10^{-7}$ attains a \emph{larger} dust concentration than $\alpha_\text{visc}=10^{-8}$.

\begin{figure*}
    \centering
    \includegraphics[width=\linewidth]{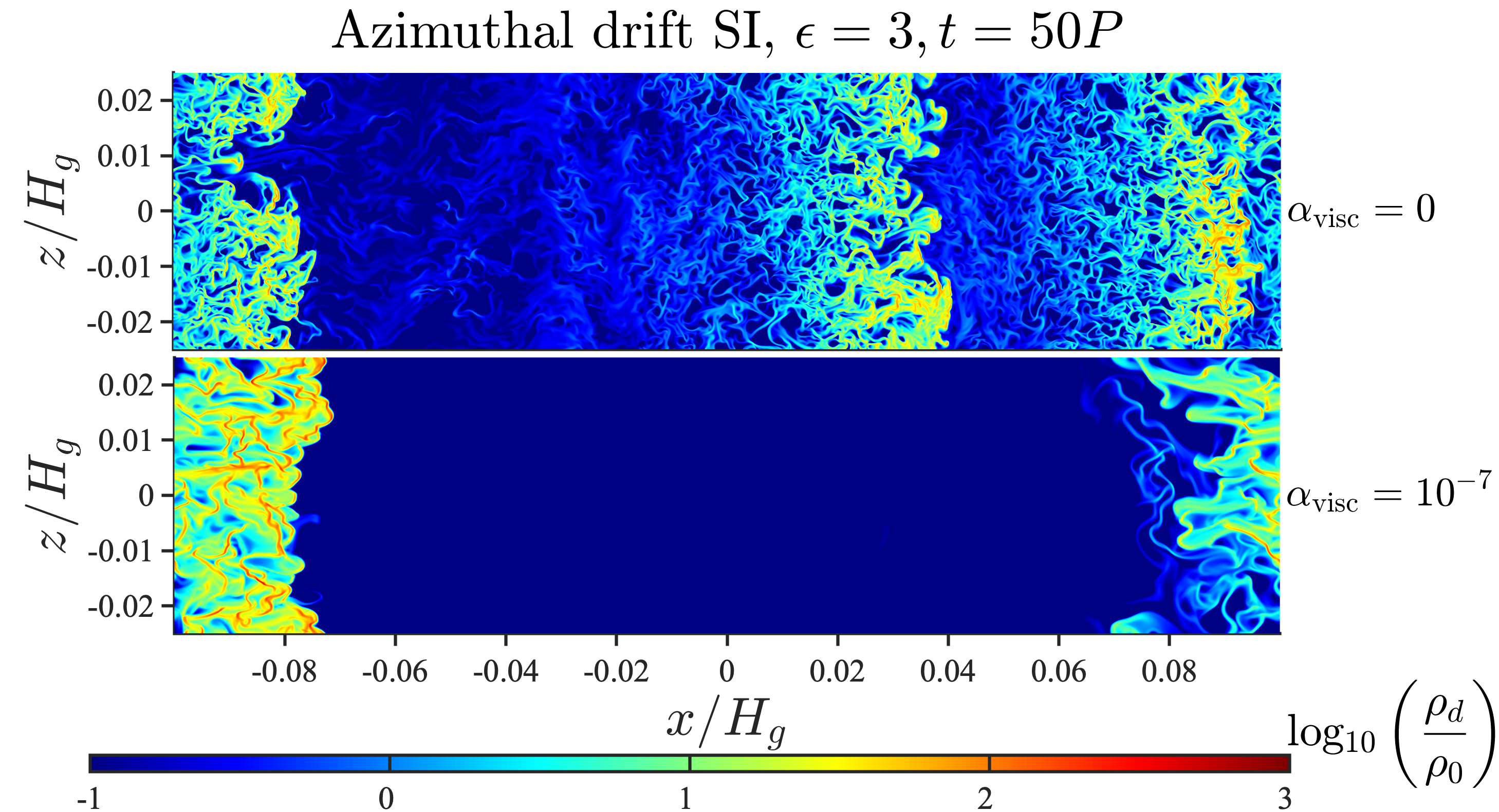}
    \caption{
    Dust density in the inviscid ($\alpha_\mathrm{visc}=0$) and viscous ($\alpha_\mathrm{visc}=10^{-7}$) runs of the AdSI ($\epsilon=3, \tilde{\eta}=0, \alpha_M=0.1$). 
    }
    \label{Paper_nvis_vs_vis_rhod3}
\end{figure*}

We find this counter-intuitive result is due to the earlier merging of filaments in the higher-viscosity case. This is shown as snapshots of the dust density in Fig. \ref{Paper_nvis_vs_vis_rhod3}, where we also show the inviscid run for comparison.  Three filaments remain in the inviscid run with small-scale eddies both within the filaments and in between them. However, the viscous run has already reached a final saturated state with a single filament. (A single filament end state was also found for $\alpha_\mathrm{visc}=10^{-8}$.) Notice the smooth flow exterior to the filament, and the larger-scale disturbances within the filament compared to the inviscid run. This suggests that small-scale disturbances work against merging, so its removal by viscosity helps clumping, at least for dust-rich disks. 

\begin{figure*}
    \centering
    \includegraphics[width=\linewidth]{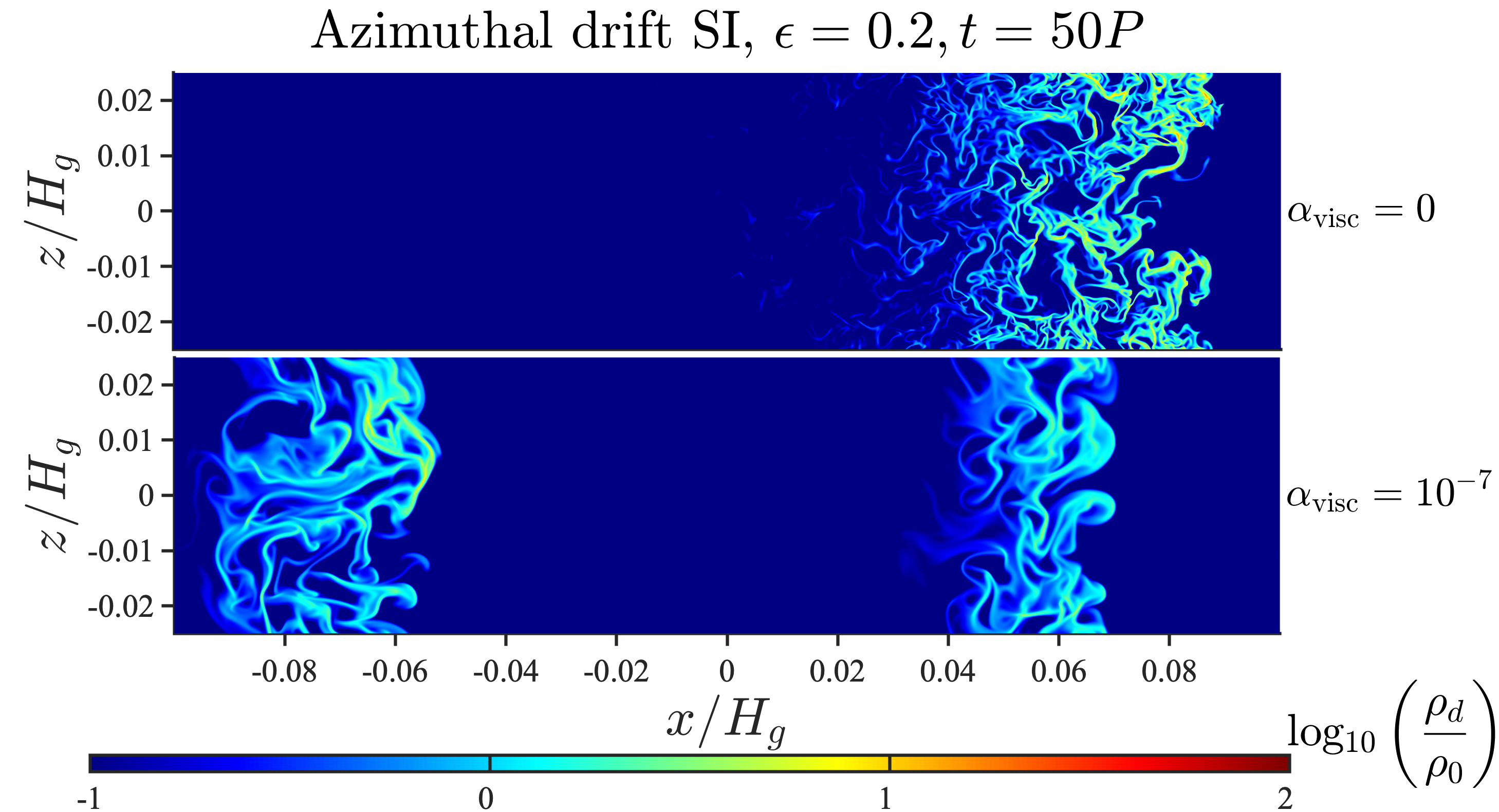}
    \caption{
    Same as Fig. \ref{Paper_nvis_vs_vis_rhod3}, but for $\epsilon=0.2$. 
    }
    \label{Paper_nvis_vs_vis_rhod02}
\end{figure*}

For $\epsilon=0.2$, dust concentrations { with $\alpha_\mathrm{visc}=10^{-7}$ are consistently weaker than with $\alpha_\mathrm{visc}=10^{-8}$, with $\rho_\text{d,max} \sim 4\rho_0$ and $14\rho_0$, respectively, at the end of the runs.} We again find this is related to filament merging. Fig. \ref{Paper_nvis_vs_vis_rhod02} shows the final dust density snapshots of the inviscid and viscous runs with $\epsilon=0.2$. In the former case, the system has already merged into a single filament, while two filaments are sustained in the latter case. Here, there is a lack of small-scale activity exterior to the filaments in either case. {It is possible that in dust-poor disks, viscosity somehow works against merging, unlike the dust-rich case. 

However, given the turbulent nature of these simulations, whether one or two filaments are formed in the end could be random. A statistical approach may necessary to assess the relation, if any, between viscosity and filament merging. 
}


\section{Discussion} \label{discussion}










\subsection{Enhanced dust concentrations in accreting disks} \label{discussion_torq}

  Our simulations demonstrate that a background gas accretion flow enhances the SI. As shown in Figs. \ref{Paper_allgrowthrates}, at fixed radial pressure gradients, $\etatilde$, the SI attains larger dust density perturbations with increasing magnitude of the accretion flow, which is parameterized by $\alpha_M$ in our models. Gas accretion becomes more important for smaller $\etatilde$. In particular, even in the absence of a radial pressure gradient ($\etatilde\equiv0$), the system is unstable for $\alpha_M>0$ and can reach strong dust clumping for sufficiently large $\alpha_M$. In the limit of $|\alpha_M/\etatilde|\gg 1$, the SI is powered by the azimuthal drift between dust and gas, unlike the classic SI, which is powered by radial drift.

  We find the AdSI can effectively concentrate dust even when the initial dust-to-gas ratio is less than unity. In our simulation with $\epsilon=0.2$, $\etatilde=0$, and $\alpha_M=0.1$, the AdSI led to an $O(10^2)$ times increase in $\rhod$. While the saturated $\epsilon\sim { 44}$ is insufficient for gravitational collapse (unless the disk is somewhat massive with a Toomre $Q\lesssim 10$), it does mean that dust feedback becomes dynamically important via the AdSI. This result is distinct from the classic SI. For example, \cite{johansen07}'s `AA' simulation with $\epsilon=0.2$\footnote{\citeauthor{johansen07}'s `AA' run has the same physical parameters as our E02eta005am0 simulation, but we were not able to run it to saturation at our grid resolutions due to computational cost.} only attains a $\sim 20\%$ increase in $\epsilon$ in the saturated state. 
  
  On the other hand, in dust-poor disks, the AdSI is weakened by the background radial pressure gradient. We thus conclude that in dust-poor, accreting disks the SI is only relevant in regions of weak pressure gradients, and takes the form of the AdSI. 
 
  Finally, we remark that although our disk models are originally motivated by PPDs subject to magnetic torques or winds, our results are not limited to this scenario. As emphasized in \citetalias{MK22}, the key ingredient is a laminar gas accretion flow, so our findings also apply to gas accretion driven by other means. 

\subsection{Merging filaments in accreting disks} \label{discussion_filament}

We find the AdSI initially forms multiple narrowly-separated, vertically-elongated filaments, which is visible in direct snapshots of the dust density (e.g. Fig. \ref{Rhod3_in_space_0_01}). This contrasts with the classic SI where a few filaments form that is only visible in space-time plots after a vertical average (e.g. Fig. \ref{transport_classical}). 

The fact that the AdSI forms more filaments than the classic SI is qualitatively consistent with the linear theory developed in \citetalias{MK22}: as $\etatilde\to0$, unstable modes shift to larger $K_x$. For $\epsilon=3$, $\etatilde=0.05$ and $\st=0.1$, the classic SI has an optimum $K_x\sim 10^3$. However, AdSI growth rates diverge with $K_x$ unless dissipation is included. We expect a finite grid resolution to have a similar effect, implying the most unstable, resolvable AdSI has $K_x\sim10^4$. We may thus naively expect the AdSI to produce ten times the number of filaments than the classic SI, as roughly observed when comparing Figs. \ref{transport_azimuthal} and \ref{transport_classical}. 

The two filaments that emerge in our classic SI run do not merge and remain separated by $\sim 0.1\Hgas$. The lack of merging may be due to the limited integration time, domain size, or both. As shown by \cite{yang14}, with a sufficiently large horizontal box size ($1.6\Hgas$ in their case), classic SI filaments are typically separated by $0.2\Hgas$.

By contrast, we find that filaments readily merge in the AdSI run. For $\epsilon=3$, three filaments remain at the end of the simulation with a maximum separation of $0.1\Hgas$ (upper panel of Fig. \ref{Paper_nvis_vs_vis_rhod3}). We suspect this represents typical separations between AdSI filaments, since our box size is relatively large compared to AdSI radial lengthscales, and that multiple merging events have already occurred. That is, at high dust-to-gas ratios AdSI filaments are more closely packed than that produced by the classic SI. On the other hand, for $\epsilon=0.2$ only one filament remains (upper panel of Fig. \ref{Paper_nvis_vs_vis_rhod02}), in which case one cannot determine the filament separation \citep{yang14}. 

We remark that for the AdSI, it is the merging of filaments that successively increases the maximum dust-to-gas ratio. This differs from the classic SI, where $\epsilon$ quickly saturates near its final value. This can be seen in Fig. \ref{Paper_allgrowthrates} by comparing the red curve in the leftmost panel (AdSI), which shows a secular increase after the linear phase, and the green curve in the rightmost panel (classic SI), which immediately surpasses the clumping condition. It, therefore, takes longer for the AdSI to reach strong clumping than the classic SI. 

\subsection{Dust diffusion in accreting disks} \label{discussion_diffusion}

We find that when $\alpha_M>0$, i.e. with an accretion flow, azimuthal mass diffusion can be significantly stronger than in the radial and vertical directions. The classic SI, on the other hand, produces more isotropic diffusion. Taken at face value, this would suggest that in accreting disks it is more difficult for gravitational collapse to proceed in the azimuthal direction, which would favor the formation of dust rings. However, full 3D simulations are needed to address this issue. 

We remark that vertical diffusion driven by the AdSI is due to high-$K_z$ modes of  instability. Because AdSI growth rates are almost independent of $K_z$ \citepalias{MK22}, all vertical wavenumbers grow equally, but low $K_z$ modes have little vertical velocities. We caution that although AdSI can operate in razor-thin disks, such models would be misleading because they artificially suppress $K_z\neq0$ modes, which are just as important as $K_z=0$. 

 In a stratified disk, vertical diffusion balances dust settling to set the particle layer scale height, 
 \begin{align}
 \Hd\simeq \sqrt{\frac{\alpha_{g,z}}{\st}}\Hgas, \label{Hdust}
 \end{align}
 { for $\alpha\ll\st$} \citep{dubruelle95}.  We find $\alpha_{g,z}$ is $O(10^{-6})$ for the classic SI and the AdSI, which gives $\Hd\simeq 0.003\Hgas$ in both cases for $\st=0.1$. One therefore cannot distinguish between the AdSI and the classic SI from the particle scale height alone. 
 
 However, when both a radial pressure gradient and a strong torque are present, for example in our run with $\epsilon=3, \etatilde=0.05$, and $\alpha_M=0.1$, $\alpha_{g,z}$ can reach $O(10^{-4})$, giving $\Hd\simeq 0.03\Hg$. Whether this is the classic SI enhanced by an accretion flow or the simultaneous presence of the classic SI and the AdSI, should be clarified. 
 
 In any case, given that most disk regions possess a non-zero radial pressure gradient (with a nominal $\etatilde\simeq 0.05$), our results suggest that in parts of the disk with rapid gas accretion, dust layers can be puffed-up to a few percent of the gas scale height just from the SI. 
 
\subsection{Planetestimal formation via the AdSI} \label{discussion_realPPDs}

Previous numerical simulations of stratified disks show that dust clumping via the classic SI is favored by weaker radial pressure gradients \citep[smaller $\etatilde$, ][]{bai10c,sekiya18} at fixed metallicities (the vertically-integrated dust-to-gas ratio). However, this result cannot be extrapolated to the limit of $\etatilde\to 0$, for example near a pressure bump, since the linear instability shifts to arbitrarily small scales and is stabilized when $\etatilde=0$. 

The AdSI does not require $\etatilde\neq0$, provided that the gas undergoes accretion. However, at dust-to-gas ratios $\gtrsim 1$, the AdSI takes longer to produce strong clumping than the classic SI, because the AdSI requires dust filaments to merge: the system evolves through a series of quasi-steady states separated by merging events that increase the maximum dust densities.  

Furthermore, a dust clump, even at the Roche density, must still exceed a critical lengthscale $l_c =  \frac{1}{3}\sqrt{\delta/\st}\Hgas$ to overcome diffusion and undergo gravitational collapse \citep{klahr20}. Here, $\delta$ is a dimensionless measure of dust diffusion due to turbulent gas stirring. In our fiducial AdSI simulation (E3eta0am01), $\delta$ is dominated by azimuthal diffusion with $\delta\sim 10^{-5}$; while for the classic SI (E3eta005am0) $\delta\sim  10^{-6}$. This implies that AdSI-clumps should be $\sim 3$ times larger than classic SI-clumps to collapse. 

We suggest that, while in disk regions of vanishing pressure gradients the AdSI can develop (whereas the linear, classic SI cannot), planetesimal formation via the AdSI would still be less efficient than it would be via the classic SI in regions of nominal pressure gradients. 


The above discussion is based on our simulations with $\epsilon=3$. However, another key distinction between the AdSI and the classic SI is that the AdSI can raise the local dust-to-gas ratio to $\gtrsim O(1)$ even if $\epsilon<1$ initially; while none of our classic SI runs initialized with $\epsilon=0.2$ attain order-unity dust-to-gas ratios. This suggests a mechanism of `AdSI-assisted' planetesimal formation via the classic SI in low metallicity disks, as follows. 

Consider an accreting disk with a weak, but non-vanishing radial pressure gradient and $\epsilon<1$. The AdSI first develops and increases $\epsilon>1$, from which the classic SI can then develop and drive strong clumping, provided it can overcome the underlying, small-scale AdSI turbulence \citep{chen20,umurhan20,gole20}. In this picture, the AdSI provides a mechanism to raise the local metallicity, as required for planetesimal formation via the classic SI \citep{johansen09}.

\subsection{Caveats and outlook}



As discussed above, our fiducial classic SI and AdSI simulations imply dust layer thicknesses of about $2\Hd\simeq 0.06\Hgas$, which is comparable to our vertical domain size $L_z$. For less unstable runs, $2\Hd$ can be significantly smaller than $L_z$, implying stratification effects could be significant. 

Furthermore, the AdSI produces vertically-extended filaments. It is not clear if these are geometrically compatible with a background vertical disk structure. It will therefore be necessary to extend the current models to stratified disks. Additional complications are expected, however, for example from the vertical shear in the disk's rotation  \citep{ishitsu09,lin21}.

In addition, our models impose axisymmetry and neglect particle self-gravity, which prohibits a proper assessment of planetesimal formation. Although some of our runs do meet the condition for strong clumping, whether or not gravitational collapse will follow can only be determined with full 3D, stratified models with self-gravity. This is particularly important for the AdSI as it appears to be anisotropic with more efficient azimuthal diffusion.

We model accretion mediated by a global magnetic field by applying a constant gas torque in the shearing box. In reality, this torque results from non-ideal MHD effects \citep{lesur20} that likely vary with space, time, and the gas-dust dynamics. Our hydrodynamic approach automatically eliminates genuine MHD phenomena that may also interact with dust dynamics \citepalias{MK22}. Global MHD simulations including dust and feedback are therefore necessary to verify our main results.

{ 
We mostly considered inviscid disks. Even our most viscous run with 
$\alpha_\mathrm{visc}=10^{-6}$ is still significantly smaller than what might be expected in reality. In PPDs, hydrodynamic instabilities produce $\alpha_\mathrm{SS}\gtrsim 10^{-5}$ \citep{lesur22}, which may or may not translate to an equivalent $\alpha_\mathrm{visc}$. Nevertheless, \emph{if} hydrodynamic turbulence behaves like a Navier-Stokes viscosity, then the AdSI can be expected to grow much more slowly than in the cases examined here, if at all. Whether or not the AdSI can produce significant dust concentrations at $\alpha_\mathrm{visc}\gtrsim 10^{-6}$ will need to be explored with long-term simulations.  
}

Finally, we have only considered one species of dust grains characterized by a single stopping time. Recent work has shown that the classic SI can be weakened when a particle size distribution is considered such that the total dust-to-gas ratio and maximum Stoke number are both $\lesssim 1$ \citep{krapp19,zhu20,paardekooper20}. When either of these is overcome, the polydisperse SI grows fast and behaves similarly to the single-species SI \citep{yang21}. In light of this, and since the AdSI can grow rapidly for both high and low dust-to-gas ratios, it may be more robust to a particle size distribution. This will need to be verified or refuted with multi-species simulations of the AdSI.

\section{Summary} \label{summary}

In this paper, we conduct high-resolution, axisymmetric, unstratified shearing box simulations of a dusty PPD with an underlying gas accretion flow. We are motivated by recent MHD simulations of PPDs that exhibit laminar gas accretion driven by magnetic winds and stresses. We are interested in how the SI operates in such an accreting disk. As a simplification, we forgo a full MHD treatment and instead apply a torque onto the gas in an otherwise hydrodynamic model. 

We previously demonstrated, using linear theory, a modified form of the SI in accreting disks, the AdSI, that is driven by the azimuthal drift between dust and gas \citepalias{MK22}. The AdSI is unlike the classic SI of \cite{youdin05} that is driven by the radial drift between dust and gas. Consequently, the AdSI can operate in the absence of a radial pressure gradient, which is a prerequisite for the classic SI. 

Here, we explore the nonlinear evolution of the AdSI. Our main findings are as follows.

\begin{enumerate}

   \item We verify the linear theory of the AdSI developed in \citetalias{MK22}, showing that even in the absence of a radial pressure gradient, an accreting, dusty disk can be unstable, evolve into a turbulent state, and trigger dust concentrations in vertically-extended filaments. 

   \item AdSI-induced dust filaments merge over time. For dust-rich disks initialized with a dust-to-gas ratio of 3, filament merging eventually drives the maximum dust-to-gas ratios to exceed 100, which is the critical value for the gravitational collapse of a dust clump in a disk with Toomre parameter of about 20.
   
   \item Even in dust-poor disks with an average dust-to-gas ratio of 0.2, the AdSI can concentrate dust to a maximum dust-to-gas ratio of about  ${ 40}$. This contrasts with previous studies on the classic SI in dust-poor disks, which only yield $20\%$ enhancement in dust densities \citep[e.g.][]{johansen07}.  
   
   \item AdSI-driven turbulence is anisotropic with azimuthal mass diffusion coefficients up to an order of magnitude larger than that in the radial and vertical directions.  
   
\end{enumerate}

We speculate that planetesimal formation directly via the AdSI in the absence of a radial pressure gradient is still less efficient than that via the classic SI in the presence of a non-vanishing radial pressure gradient. This is because AdSI-driven disks only gradually develop dense dust clumps via filament merging, whereas the classic SI quickly attains it. However, in disk regions of low metallicity, the AdSI can still raise midplane dust-to-gas ratios to values above unity, which may then enable the classic SI to ensue more effectively and facilitate planetesimal formation.

 \begin{acknowledgements}
 {We thank the anonymous reviewer for a prompt report and helpful comments and suggestions.} This work is supported by the National Science and Technology Council (grants 107-2112-M-001-043-MY3, 110-2112-M-001-034-, 111-2112-M-001-062-, 110-2124-M-002-012-, 111-2124-M-002-013-) and an Academia Sinica Career Development Award (AS-CDA-110-M06). Simulations were carried out at the TIARA clusters at ASIAA and the Taiwania-3 cluster at the National Center for High-performance Computing (NCHC) of the National Applied Research Laboratories (NARLabs) in Taiwan. We thank NCHC and NARLabs for providing computational and storage resources. C-YH is supported in part by NASA 80NSSC18K1095 at the University of Virginia.
\end{acknowledgements}

\appendix \label{appendix}

\section{Code test}\label{code_test}

We test our modified \textsc{fargo3d} code by simulating the AdSI and comparing growth rates with that from the linear theory developed by \citetalias{MK22}. The equilibrium state is the same as that used in the main text (\S\ref{eqm_state}). Here, we set $\etatilde\equiv 0$ so the classic SI does not apply. We fix $\epsilon=3$ and consider cases with and without a gas viscosity and a corresponding dust diffusion. In viscous runs we use $\alpha_\mathrm{visc}=10^{-8}$. 

We seed the initial conditions with unstable modes obtained from solving the corresponding linear eigenvalue problem described in \citetalias{MK22} (Appendix B, but without magnetic fields). We reproduce them here for convenience:
\begin{align} 
&\sigma\Wpert = -\ii k_x v_x \Wpert  - \ii k_{x} \delta v_x - \ii k_{z} \delta v_{z}, \label{lin_gas_mass}\\ 
&\sigma \delta v_x = -\ii k_x v_x \delta v_x + 2 \Omega \delta v_y - \ii k_x C_s^2 \Wpert - \frac{\epsilon}{\taus}(w_{x}-v_{x})\left(\Wpert - \Qpert\right)+ \frac{\epsilon}{\taus}(\delta w_x - \dd v_x) + \dd F_x^{\rm visc} , \label{lin_gas_mom_x}\\
&\sigma \delta v_y = -\ii k_x v_{x} \delta v_y - \frac{\Omega}{2} \delta v_x - \frac{\epsilon}{\taus}(w_{y}-v_{y}) \left(\Wpert-\Qpert\right) + \frac{\epsilon}{\taus} (\delta w_y - \delta v_y) + \dd F_y^{\rm visc},  \label{lin_gas_mom_y}\\
&\sigma \delta v_z = -\ii k_x v_{x} \delta v_z   + \frac{\epsilon}{\taus}(\delta w_z - \delta v_z) - \ii k_z C_s^2 \Wpert + \dd F_z^{\rm visc}, \label{lin_gas_mom_z}\\
&\sigma \Qpert = -\ii k_x w_{x} \Qpert -\ii k_x \delta w_x -\ii k_z\delta w_z - D k^2 \left(\Qpert-\Wpert\right), \label{lin_dust_mass}\\
&\sigma \delta w_x = -\ii k_x w_{x} \delta w_x + 2 \Omega \delta w_y  - \frac{1}{\taus} (\delta w_x - \delta v_x),\label{lin_dust_mom_x}\\
&\sigma \delta w_y = -\ii k_x w_{x} \delta w_y - \frac{\Omega}{2} \delta w_x  - \frac{1}{\taus} (\delta w_y - \dd v_y), \label{lin_dust_mom_y}\\
&\sigma \delta w_z = -\ii k_x w_{x} \delta w_z - \frac{1}{\taus} (\delta w_z - \delta v_z), \label{lin_dust_mom_z}
\end{align}
where the linearized viscous forces\footnote{The viscous terms in \citetalias{MK22}'s Eqs. B2-B4 were erroneously written to correspond to a viscous stress tensor of the form $\nu\nabla^2\bm{v}$. The actual form used in that and the present work is given by Eq. \ref{viscous_stress} and its linearized version is given by Eqs. \ref{lin_visc_x}---\ref{lin_visc_z}. However, since the gas dynamics are almost incompressible, whether $\nu\nabla^2\bm{v}$ or Eq. \ref{viscous_stress} is used makes little difference.} are
\begin{align}
    \dd F_x^{\rm visc} &=  - \nu \left[ \left(\frac{4}{3}k_x^2+k_z^2 \right)\dd v_x  -\frac{1}{3} k_x k_z \dd v_z \right], \label{lin_visc_x}\\
    \dd F_y^{\rm visc} &=  - \nu  k^2\dd v_y  , \label{lin_visc_y}\\
    \dd F_z^{\rm visc} &=  - \nu \left[ \left(k_x^2+\frac{4}{3}k_z^2 \right)\dd v_z  -\frac{1}{3} k_x k_z \dd v_x \right]. \label{lin_visc_z}
\end{align}
In the above, $\delta A$ is a complex perturbation amplitude of any field $A$; $k_{x}$ and $k_{z}$ are real radial and vertical wavenumbers, respectively, with $k^2 \equiv k_x^2 + k_z^2$; and the complex growth rate $\sigma = s - \ii \omega$, where $s$ is the real growth rate and $\omega$ is the oscillation frequency. 

We normalize the wavenumbers by $\Hg$ so that $k_{x,z}\equiv K_{x,z}/\Hgas$. We follow \cite{youdin07} and add a pair of eigenmodes with oppositely-signed $K_z$ to produce standing waves in $z$. 

For the tests below we fix $K_z=100$. The computational domain is one wavelength in each direction, i.e. $L_{x,z}=2\pi/k_{x,z}$, and we adopt a resolution of $N_{x}\times N_z=256\times256$.  We measure growth rates by tracking the evolution in the maximum dust density perturbation. 

\begin{figure}
    \centering
    \includegraphics[width=\linewidth]{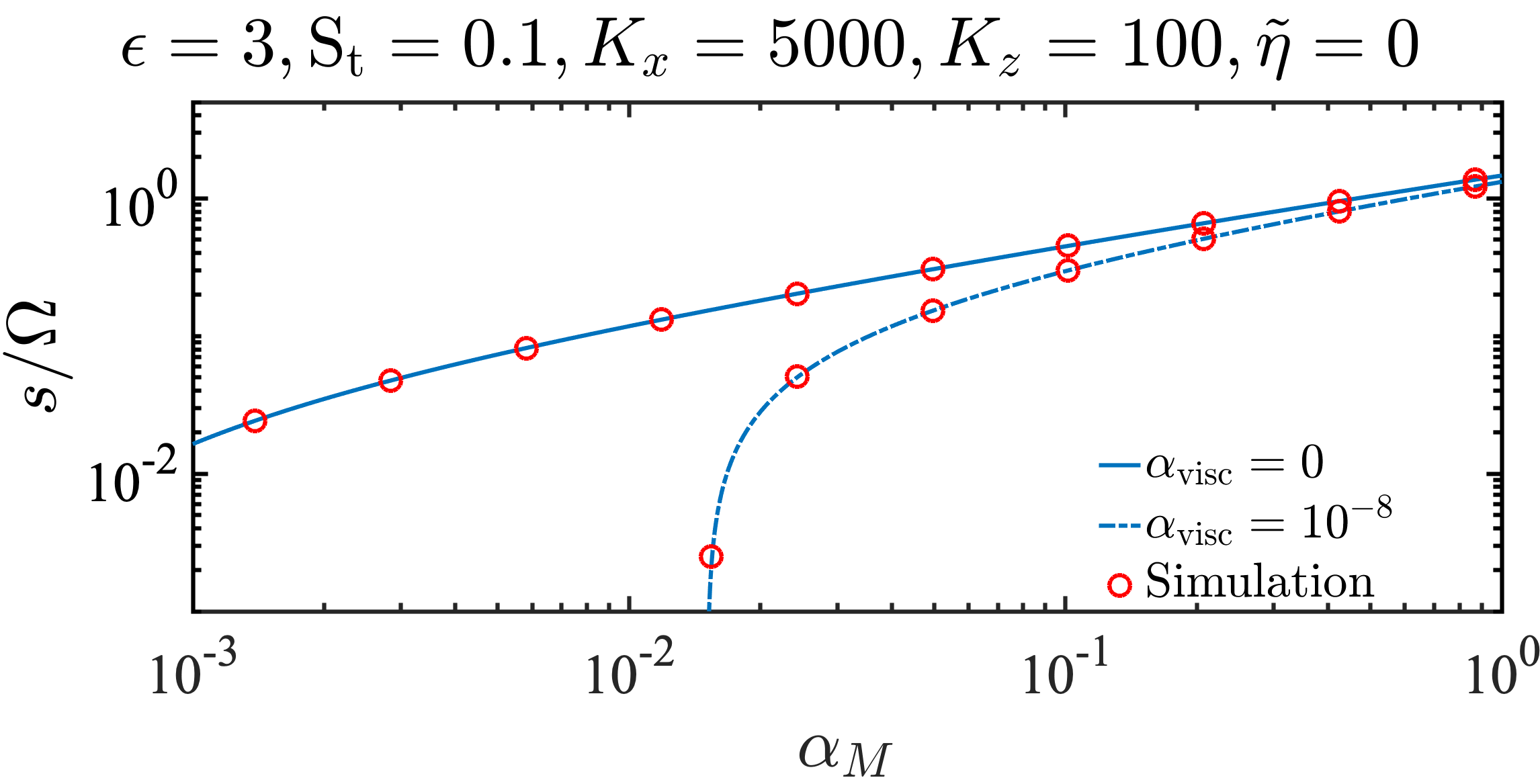}
    \caption{
    Comparison between growth rates of the AdSI computed from linear theory (blue curves) and that measured from numerical simulations (red circles), for fixed $\epsilon=3$, $\st=0.1$, $\etatilde=0$, $K_z=100$, as a function of $\alpha_M$.
    }
    \label{Appendix_CT}
\end{figure}

\begin{figure}
    \centering
    \includegraphics[width=\linewidth]{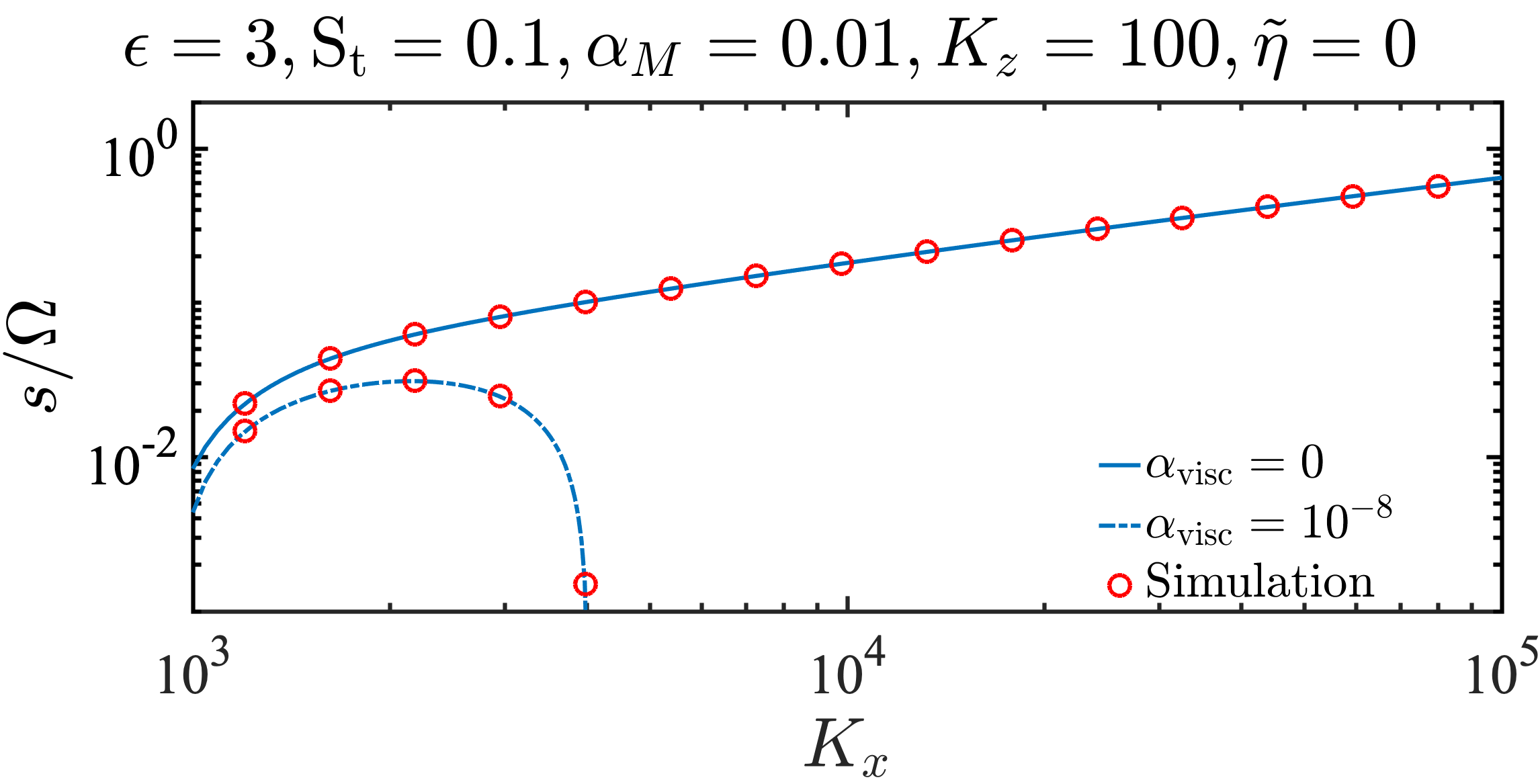}
    \caption{Similar to Fig. \protect\ref{Appendix_CT} but for fixed $\alpha_M=0.1$ and varying $K_x$. 
    }
    \label{Appendix_CT_kx}
\end{figure}

Fig. \ref{Appendix_CT} show growth rates as a function of the Maxwell stress applied to the gas, $\alpha_M$, for fixed $K_x=5000$; while Fig. \ref{Appendix_CT_kx} show growth rates as a function of $K_x$ for fixed $\alpha_M=0.1$. Simulation results are shown in red circles and linear theory is shown as the blue curves. All of the measured growth rates from the numerical simulations are in excellent agreement with the theoretical values. 

We note that the linear AdSI does not depend on vertical direction \citepalias[][except at high $K_z$ where viscosity and diffusion take effect, if included]{MK22}, i.e.
the instability persists for $K_z = 0$. We have verified this by performing one-dimensional simulations (by setting $N_z=1$) and found the same growth rates as in the ones above and in agreement with linear theory. 

\bibliographystyle{aasjournal}
\bibliography{ref}

\end{document}